\documentstyle[11pt]{article}

\makeatletter
\@addtoreset{equation}{section}
\makeatother

\def\+{\;+\;}
\def\-{\;-\;}
\def\*{\,\cdot\,}
\def\u{\mbox{\bf\sf u}}
\def\w{\mbox{\bf\sf w}}
\def\s{\mbox{\bf\sf s}}
\def\v{\mbox{\bf\sf v}}
\def\x{\mbox{\bf\sf x}}
\def\y{\mbox{\bf\sf y}}
\def\z{\mbox{\bf\sf z}}
\def\A{\mbox{\bf\sf A}}

\def\C{\mbox{\bf\sf C}}

\def\H{\mbox{\bf\sf H}}
\def\W{\mbox{\bf\sf W}}
\def\U{\mbox{\bf\sf U}}

\def\a{\mbox{\bf\sf a}}
\def\b{\mbox{\bf\sf b}}
\def\c{\mbox{\bf\sf c}}
\def\d{\mbox{\bf\sf d}}

\def\f{\mbox{\bf\sf f}}
\def\ds{\displaystyle}
\def\R{\mbox{\sf R}}
\def\U{\mbox{\sf U}}
\def\P{\mbox{\sf P}}
\def\L{\mbox{\sf L}}
\def\J{\mbox{\sf J}}
\def\smb{{\displaystyle\star}}
\def\lgr{\bigtriangleup}
\def\rgr{\bigtriangledown}

\begin{document}



\title{QUANTUM 2+1 EVOLUTION MODEL}

\author{
S. M. Sergeev\\
Branch Institute for Nuclear Physics, Protvino 142284, Russia.\\
E-mail: sergeev\_ms@mx.ihep.su
}

\date{October, 1998}

\maketitle
\abstract{
A quantum evolution model in $2+1$ discrete space -- time,
connected with 3D fundamental map $\R$, is investigated. 
Map $\R$ is derived as a map providing a zero
curvature of a two dimensional lattice system called
``the current system''. In a special case of the local Weyl algebra
for dynamical variables the map appears to be canonical one and it
corresponds to known operator-valued $\R$ -- matrix.
The current system is a kind
of the linear problem for $2+1$ evolution model.
A generating function for the integrals of motion for the evolution is
derived with a help of the current system.      The subject of the
paper is rather new, and so the perspectives of further investigations
are widely discussed.
}

\bigskip
\noindent
{\em PACS:} 05.50; 02.10; 02.20.\\
{\em Mathematics Subject Classifications (1991):} 
47A60, 47A67, 22D25.\\
{\em Keywords:} Discrete space -- time evolution models;
$2+1$ integrability; Tetrahedron equation



\section{Introduction}

\subsection{3D integrable models}

In 3D integrable models the Tetrahedron equation (TE) takes place
of the Yang -- Baxter equation (YBE) in 2D. Having got a solution
of TE, one may hope to construct a 3D integrable model. In the
case of finite number of states one may construct usual layer --
to -- layer transfer matrices $T$, so that TE provides the
commutability of them \cite{zam-solution,bs-dsimpl,baxter-pf}.
Such finite states models are interpreted usually as statistical
mechanics models.
Really only one such model still exists, the Zamolodchikov --
Bazhanov -- Baxter model
\cite{zam-solution,baxter-pf,bb-first,mss-vertex}. The uniqueness
does not mean that 3D world has no interest.

When 3D $\R$-matrices have infinitely many states, which is more
usual in 3D, very natural  is to investigate  a kind of transfer
matrices that has no hidden space. We denote such transfer
matrices as $\U$ {\em versus} the notation for usual transfer
matrix $T$. Matrices $\U$ commute with the set of $T$-s, but have
no degrees of freedom when the set of $T$ is fixed. Thus $\U$
resemble a hamiltonian. Conventionally models with infinitely many
states are regarded as field theory ones.

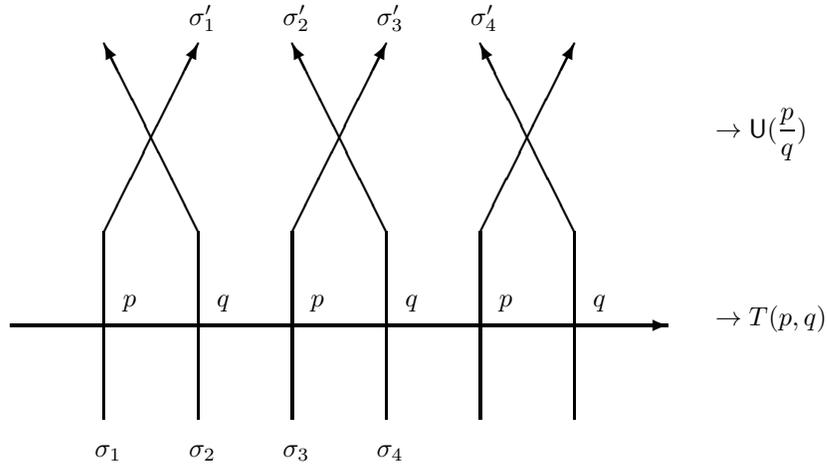
\begin{figure}
\begin{center}
\setlength{\unitlength}{0.25mm} 
\thicklines
\thicklines
\begin{picture}(450,300)
\put(0,0){
\begin{picture}(450,300)
\put(0,100){\vector(1,0){350}}
\multiput(50,50)(50,0){6}{\line(0,1){100}}
\multiput(50,150)(100,0){3}{\vector(1,2){50}}
\multiput(100,150)(100,0){3}{\vector(-1,2){50}}
\put(60,110){$p$}\put(110,110){$q$}
\put(160,110){$p$}\put(210,110){$q$}
\put(260,110){$p$}\put(310,110){$q$}
\put(45,30){$\sigma_1$}\put(95,260){$\sigma_1'$}
\put(95,30){$\sigma_2$}\put(145,260){$\sigma_2'$}
\put(145,30){$\sigma_3$}\put(195,260){$\sigma_3'$}
\put(195,30){$\sigma_4$}\put(245,260){$\sigma_4'$}
\put(375,200){$\ds \rightarrow \U({p\over q})$}
\put(375,100){$\ds \rightarrow T(p,q)$}
\end{picture}}
\end{picture}
\end{center}
\caption{Configuration $\U\*T$ in $2D$.}
\label{fig-ut}
\end{figure}

The structure of $\U$ is clarified in Fig. \ref{fig-ut} for 2D
case. Here $p$ and $q$ stand for the spectral parameters of the
vertices of $T$, $p/q$ consequently is the argument of the
vertices of $U$, and $\sigma_j$ and $\sigma_j'$ are the indices
taking values in a finite set. This $2D$ picture we give just an
example for the sake of clearness. The $1+1$ evolution models,
connected with classical or quantum bilinear Hirota or Hirota-Miwa
equitations on the lattice, are always formulated in terms of
$U$-type evolution operators, see for example
\cite{fv-hirota,fv-quantization} and references therein.

In 3D, $\U$-matrix appears as an element of a cubic lattice
included between two nearest inclined planes. We do not draw the
graphical representation of 3D $\U$ here, we will consider
sections of the cubic lattice by two, in- and out-, inclined
planes mentioned. A two dimensional lattice appearing in  such
sections is called the kagome lattice and we will consider it in
details below. The first who considered $\U$ -- matrices in $3D$,
constructed with a help of finite -- state $R$ -- matrix, and
constructed some eigenvectors for it, was I. Korepanov
\cite{korepanov-diss,korepanov-u,korepanov-BA}.

\subsection{3D integrability: usual approach}

So, the origin of 3D integrability is a solution of TE. Those who
dealt with it know that it is practically impossible to find it
directly. For example, even to prove TE analytically for an ansatz
given and tested numerically is  bloody complicated
\cite{baxter-proof,kms-stsq}. This means, we guess, there must be
an alternative way of a 3D Boltsmann weights' derivation.

Primitive way is to find a solution of TE is to consider 
the intertwining
relation for a triple sets of 2D $L$-matrices,
\begin{equation}
\ds
\sum_{j_1,j_2,j_3}\;\;
R_{i_1,i_2,i_3}^{j_1,j_2,j_3}\;
\bigl( L_{j_1}^{k_1}\bigr)_{1,2}\;
\bigl( L_{j_2}^{k_2}\bigr)_{1,3}\;
\bigl( L_{j_3}^{k_3}\bigr)_{2,3}\;=\;
\sum_{j_1,j_2,j_3}\;\;
\bigl( L_{i_3}^{j_3}\bigr)_{2,3}\;
\bigl( L_{i_2}^{j_2}\bigr)_{1,3}\;
\bigl( L_{i_1}^{j_1}\bigr)_{1,2}\;
R_{j_1,j_2,j_3}^{k_1,k_2,k_3}\;,
\end{equation}
where the structure of $L_{1,2}L_{1,3}L_{2,3}$ {\em versus}
$L_{1,2}L_{1,3}L_{2,3}$ is the Yang -- Baxter structure, and the extra
indices  correspond to the possibility to consider coefficients
$R_{i_1,i_2,i_3}^{k_1,k_2,k_3}$ as 3D $\R$ -- matrix. TE appears as the
admissibility condition for
\begin{equation}
\ds
L_{1,2}\;L_{1,3}\;L_{2,3}\;L_{1,4}\;L_{2,4}\;L_{3,4}
\;\mapsto\;
L_{3,4}\;L_{2,4}\;L_{1,4}\;L_{2,3}\;L_{1,3}\;L_{1,2}\;.
\end{equation}
Another scenario is the Zamolodchikov tetrahedral algebra
\begin{equation}
\ds
\Psi_{1,2}^{k_1}\;\Psi_{1,3}^{k_2}\;\Psi_{2,3}^{k_3}\;=\;
\sum_{j_1,j_2,j_3}\;\;
\Psi_{2,3}^{j_3}\;\Psi_{1,3}^{\j_3}\;\Psi_{1,2}^{\j_1}\;
R_{j_1,j_2,j_3}^{k_1,k_2,k_3}\;.
\end{equation}
In the compact form, introducing the formal basis for the indices
of $R$, $e(i,j)\;\equiv\;|i><j|$,
\begin{equation}
\ds R_{a,b,c}\;=\;\sum_{j,k}\;\;
R_{j_1,j_2,j_3}^{k_1,k_2,k_3}\;\; e_a(j_1,k_1)\; e_b(j_2,k_2)\;
e_c(j_3,k_3)\;.
\end{equation}
TE looks like
\begin{equation}\label{te}
\ds
\R_{1,2,3}\*\R_{1,4,5}\*\R_{2,4,6}\*\R_{3,5,6}\;=\;
\R_{3,5,6}\*\R_{2,4,6}\*\R_{1,4,5}\*\R_{1,2,3}\;,
\end{equation}
where the alphabetical indices, labelling the numbers of the spaces, are
conventionally changed to the numerical indices, such change we will make
frequently.

Most amusing thing is that all these really give a 3D $R$-matrix:
Korepanov's $R$-matrix.\ \cite{korepanov-diss,korepanov}. Korepanov's
$R$-matrix as well as Hietarinta's one are some special cases of more
general, complete $R$-matrix derived by Sergeev, Mangazeev and Stroganov
\cite{mss-vertex}, and complete $R$-matrix is equivalent to
Zamolodchikov -- Bazhanov -- Baxter's weights in the thermodynamic limit.

\subsection{3D integrability: functional approach}

A way to get something else in 3D is to refuse the finite number of
states in the previous approach. Namely, 3D models appear in the
local Yang -- Baxter equation (LYBE) approach. LYBE
(i.e. a Yang -- Baxter equation with different ``spectral'' parameters
in the left and right hand sides) can be adapted to a
discrete space -- time evolution of the triangulated 
two dimensional oriented
surface as a kind of zero curvature condition
\cite{maillet-nijhoff-talk,maillet-nijhoff,maillet}.

In few words, if a matrix $L_{i,j}(\x)$, acting as usual
in a tensor product of two
finite dimensional spaces labelled by numbers $i$ and $j$,
with some fixed functional structure and depending on a set of
parameters $\x$, obeys the equation
\begin{equation}\label{lybe}
\ds
L_{1,2}(\x_a)\; L_{1,3}(\x_b)\; L_{2,3}(\x_c)\;\;=\;\;
L_{2,3}(\x_c')\;L_{1,3}(\x_b')\;L_{1,2}(\x_a')\;,
\end{equation}
called the local Yang--Baxter equation, so that parameters
$\x_a$, $\x_b$ and $\x_c$ are independent and
$\x_a'$, $\x_b'$ and $\x_c'$ can
be restored from (\ref{lybe}) without any ambiguity,
\begin{equation}\label{fun-map}
\ds
\x_a'\;=\;f_a(\x_a,\x_b,\x_c)\;,\;\;\;
\x_b'\;=\;f_b(\x_a,\x_b,\x_c)\;,\;\;\;
\x_c'\;=\;f_c(\x_a,\x_b,\x_c)\;,
\end{equation}
then the functional map $\R$ is introduced:
\begin{equation}
\ds
\R_{a,b,c}^{}\*\varphi(\x_a,\x_b,\x_c)\*\R_{a,b,c}^{-1}\;=\;
\varphi(\x_a',\x_b',\x_c')\;\;\;\;\forall\;\;\;\varphi(...)\;.
\end{equation}
Due to the difference of the ``spectral'' parameters in the left
and right hand sides of LYBE, any shift of a line of a two
dimensional lattice, constructed with a help of $L_{i,j}(\x_{i,j})$,
changes the set of parameters $\x_{i,j}$.
Partially, if any shift of the lines can be decomposed into 
primitive shifts like (\ref{lybe}) in different ways, then 
corresponding different products of $\R$-s coincide. The 
basic example of this is the functional Tetrahedron equation.

Suppose we move all the lines of the lattice in some regular way,
conserving a structure of the lattice. Then the change of
parameters $\x_{i,j}$ can be considered as an one-step evolution
of the dynamical variables $\x_{i,j}$ governed by an appropriately
defined evolution operator $\U\;=\;\prod_{\mbox{\scriptsize
triangles}}\;\R$. This evolution is integrable due to uniqueness
of LYBE and (\ref{fun-map}). The partition function for the
lattice becomes the natural integral of motion. In terms of the
transfer matrices, the partition function is the $T$-type transfer
matrix. Being functional, these $\R$-operators correspond to
something infinitely dimensional. Contrary to the previous finite
dimensional $R$-matrices, there are known a lot of such
$\R$-operators. The reader can find an interesting set of such
simplest functional $\R$-s in \cite{oneparam}.

A quantization of known functional $\R$-s is still open problem.
Simplest functional $\R$-s are to be regarded as some functional limits
of multivariable $\R$-s with a symplectic structure conserving. The problem
is to rise  known $\R$-s to the complete phase space case, this is done
just for a couple of $\R$-s.

\subsection{3D integrability: general concept of evolution}

Here we discuss, what else can be invented to get a 3D
integrability.

The main observation is that the relations like tetrahedral
Zamolodchikov algebra and LYBE have usual graphical interpretation
as the equality of the objects assigned to two similar graphs.
These graphs are the triangles, and we will denote them briefly as
$\lgr$ and $\rgr$. Left hand side type graph $\lgr$ corresponds to
a product like $L_{1,2}\;L_{1,3}\;L_{2,3}$, and right hand side
type graph $\rgr$ corresponds to $L_{2,3}\;L_{1,3}\;L_{1,2}$.
Algebraic objects are assigned to the elements of these graphs. In
the case of LYBE these algebraical objects are matrix $L$ 
with the indices assigned to the edges (in the form of subscript, 
for $L_{1,2}$ $1$ and $2$ stand for the edges), 
and parameter $\x$ assigned to the vertex. 
From 3D point of view $\x$-s are the dynamical variables,
whereas $L$ and its indices are auxiliary objects. The equality of l.h.s.
of LYBE and r.h.s. of LYBE gives the notion of {\bf the algebraic
equivalence} of $\lgr$ and $\rgr$. Note, this form of the
algebraic equivalence is {\bf not obligatory} !

\bigskip

{\bf 3D integrability we can get from any other decent definition
of an algebraic equivalence.}

\bigskip

In this paper we consider a system, associated with a set of
equivalent planar graphs.  We propose another notion of an
algebraic equivalence of equivalent graphs.

\bigskip

We will deal with all elements of the cw-complex, so we start from
recalling the relationship between the elements of a planar graph
and repeating some definitions.

Consider a graph $G_n$ formed by $n$ straight intersecting lines.
The elements of its cw-complex are the vertices, the edges and the
sites. $G_n$ consists on $\ds N_V\;=\;{n(n-1)\over 2}$ vertices,
$\ds N_S\;=\;{(n-1)(n-2)\over 2}$ closed inner sites and $\ds
N_S^*\;=\;2 n $ outer open sites, $\ds N_E\;=\;n(n-2)$ closed
inner edges and $\ds N_E^*\;=\; 2 n$ outer edges. If two graphs
$G_n$ and $G_n'$ have the same outer structure, i.e. $G_n'$ can be
obtained from $G_n$ by appropriate shift of the lines, then call
$G_n'$ and $G_n$ equivalent.

\bigskip

Suppose we assign to the elements of a graph some elementary
algebraic (maybe, the term ``arithmetical'' is more
exact) objects.  These objects are divided into two classes:
dynamical variables and auxiliary objects (see the interpretation
of $L_{1,2}(\x)$ above in this subsection). Dynamical variables
are parameters of graph $G_n$, and auxiliary objects give some two
dimensional rules of a game (like the summation over all
intermediate indices in the product of $L$-s).
Dynamical variables {\em plus} a rule of game give an algebraic
object corresponding to the whole graph (like the partition
function for $L$-s). This algebraic (arithmetical) object for
whole $G_n$ we call the observable object. Denote it $O(G_n)$. It
depends on the set of the dynamical variables.

Consider now two equivalent graphs, $G_n$ and $G_n'$. 
The algebraic problem of the equivalence arises,
\begin{equation}\label{oo}
\ds O(G_n)\;=\;O(G_n')\;.
\end{equation}
If, according to the two dimensional rules of the game, we can
get (\ref{oo}) choosing the dynamical variables for $G_n'$
appropriately for the variables of $G_n$ given, then the algebraic
equivalence makes a sense. 
If, moreover, parameters of $G_n'$ can be
restored from the algebraic equivalence condition (\ref{oo})
without any ambiguity, 
then this equivalence is decent and the integrability is
undoubted.

The algebraic equivalence usually called zero curvature, and
LYBE as the zero curvature condition as well as functional
evolution models was considered in
\cite{maillet-nijhoff-talk,maillet-nijhoff,maillet}.
Another formulation of the algebraic equivalence, different to the
LYBE approach, is Korepanov's matrix model
(see \cite{korepanov-diss,kks-fte}
and references therein). 
The formulation of the matrix model differs from
the usual assigning the vertex -- type Boltsmann 
weights to the vertices of a lattice, but functional 
evolution models probably are the same.

{\bf We chose another way.}

The method we use was formulated originally
in \cite{electric}, the classical (i.e. functional)
evolution model was described in \cite{3d-symp},
and the quasiclassical case was investigated in
\cite{ks-boson}. This paper contains the overview of the method,
and the description of the quantum evolution model. 
The main new result is the generating function for 
integrals of motion for this evolution.


\section{Auxiliary Linear Problem}

In this section we give some rules allowing one to assign
an algebraic system to a graph. The elements to which we
assign something are vertices and sites. First, we give
most general rules, which do not give an algebraic
equivalence of
equivalent graphs in general, due to a sort of a ``gauge
ambiguity''.
As a special case we find the rules
which contain not a gauge ambiguity, and so a notion of an
algebraic equivalence will be introduced.
Then we describe the
map of the dynamical variables given by the equivalence
of 2-simplices, and discuss other similar approaches giving
this map.

\subsection{General approach}

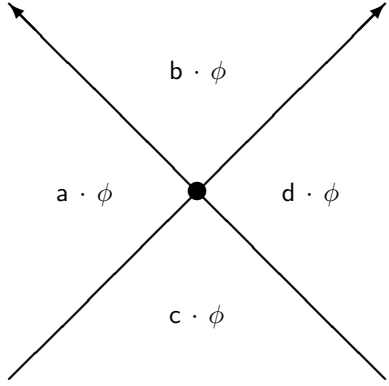
\begin{figure}
\begin{center}
\setlength{\unitlength}{0.25mm} 
\thicklines
\begin{picture}(450,200)
\put(125,0){
\begin{picture}(200,200)
\put(   0 ,   0 ){\vector( 1,1){200}}
\put( 200 ,   0 ){\vector(-1,1){200}}
\put( 100 , 100 ){\circle*{10}}
\put(  85 , 160 ){$\b\*\phi$}
\put(  85 ,  30 ){$\c\*\phi$}
\put(  25 ,  95 ){$\a\*\phi$}
\put( 145 ,  95 ){$\d\*\phi$}
\end{picture}}
\end{picture}
\end{center}
\caption{The current vertex.}
\label{fig-abcd-currents}
\end{figure}

Choose as a game the following rules:
\begin{itemize}
\item Assign to each oriented vertex $V$ an auxiliary
``internal current'' $\phi$.
Suppose this current produces four ``site currents'' 
flowing from the vertex into four adjacent faces, and
proportional to the internal current
with some coefficients $\a,\b,\c,\d$, called the 
dynamical variables,
as it is shown in Fig.
\ref{fig-abcd-currents}. All this variables,
$\phi$ and $\a,...,\d$ for different vertices are
independent for a while.
\item  Define the complete site current as an
algebraic sum of the contributions of
vertices surrounding this site.
\item  For any closed site of a lattice let 
its complete current is
zero. Such zero relations we regard as the linear
equations for the internal currents.
\item  For any graph $G_n$  the site currents assigned to
outer (open) sites we call the ``observable currents''. In part,
two equivalent graphs $G_n$ and $G_n'$ must have the same
observable currents -- this is the algebraic meaning of 
the equivalence.
\end{itemize}

Clarify these rules on the example of equivalence of $G_3$.
As it was mentioned,
this is usual Yang -- Baxter equivalence graphically,
$\lgr=\rgr$, shown in Fig. \ref{fig-YBE}.
Assign to the vertices $W_j$ of the left hand side graph
$\lgr$
the currents $\phi_j$ and the dynamical variables 
$\a_j,\b_j,\c_j,\d_j$,
and to the vertices $W_j'$ of the right hand side graph 
$\rgr$ --
the currents $\phi_j'$ and the dynamical variables
$\a_j',\b_j',\c_j',\d_j'$.
Six currents of outer sites denote as $\phi_b,...,\phi_g$,
and two zero valued currents of closed sites -- as 
$\phi_h$ and $\phi_a$
as it is shown in Fig. \ref{fig-YBE}.
Then, using the rules described above, 
we obtain the following system
of eight linear (with respect to the currents) relations:

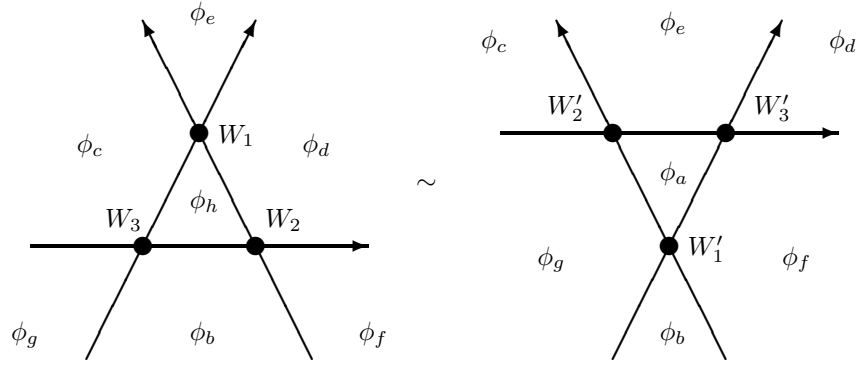
\begin{figure}
\begin{center}
\setlength{\unitlength}{0.25mm} 
\thicklines
\begin{picture}(450,200)
\put(00,00){
\begin{picture}(200,200)
\put(  10 ,  70 ){\vector(1,0){180}}
\put(  40 ,  10 ){\vector(1,2){90}}
\put( 160 ,  10 ){\vector(-1,2){90}}
\put(  70 ,  70 ){\circle*{10}}\put(50,80){$W_3$}
\put( 130 ,  70 ){\circle*{10}}\put(135,80){$W_2$}
\put( 100 , 130 ){\circle*{10}}\put(110,125){$W_1$}
\put(  95 ,  90 ){$\phi_h$}
\put(  95 , 190 ){$\phi_e$}
\put(  95 ,  20 ){$\phi_b$}
\put(  35 , 120 ){$\phi_c$}
\put( 155 , 120 ){$\phi_d$}
\put(   0 ,  20 ){$\phi_g$}
\put( 185 ,  20 ){$\phi_f$}
\end{picture}}
\put(250,00){
\begin{picture}(200,200)
\put(  10 , 130 ){\vector(1,0){180}}
\put(  70 ,  10 ){\vector(1,2){90}}
\put( 130 ,  10 ){\vector(-1,2){90}}
\put(  70 , 130 ){\circle*{10}}\put(35,140){$W'_2$}
\put( 130 , 130 ){\circle*{10}}\put(145,140){$W'_3$}
\put( 100 ,  70 ){\circle*{10}}\put(110,65){$W'_1$}
\put(  95 , 105 ){$\phi_a$}
\put(  95 , 185 ){$\phi_e$}
\put(  95 ,  20 ){$\phi_b$}
\put(   0 , 175 ){$\phi_c$}
\put( 185 , 175 ){$\phi_d$}
\put(  30 ,  60 ){$\phi_g$}
\put( 160 ,  60 ){$\phi_f$}
\end{picture}}
\put(210,90){\begin{picture}(30,50)
\put(10,10){$\sim $}
\end{picture}}
\end{picture}
\end{center}
\caption{The Yang-Baxter equivalence.}
\label{fig-YBE}
\end{figure}

\begin{equation}\label{phi-h}
\ds \phi_h\;\equiv\;\c_1^{}\*\phi_1^{}\;+\;\a_2^{}\*\phi_2^{}\;+\;
\b_3^{}\*\phi_3^{}\;=\;0\;,
\end{equation}
\begin{equation}\label{phi-bcd}
\ds\left\{\begin{array}{ccccc}
\ds\phi_b & \equiv &\ds\c_1'\*\phi_1' & = &
\ds\c_2^{}\*\phi_2\;+\;\d_3^{}\*\phi_3^{}\;,\\
&&\\
\ds\phi_c & \equiv &\ds\a_2'\*\phi_2' & = &
\ds\a_1^{}\*\phi_1\;+\;\a_3^{}\*\phi_3^{}\;,\\
&&\\
\ds\phi_d & \equiv &\ds\b_3'\*\phi_3' & = &
\ds\d_1^{}\*\phi_1\;+\;\b_2^{}\*\phi_2^{}\;,
\end{array}\right.\end{equation}
\begin{equation}\label{phi-efg}
\ds\left\{\begin{array}{ccccc}
\ds\phi_e & \equiv &
\ds\b_2'\*\phi_2'\;+\;\a_3'\*\phi_3' & = & \b_1^{}\*\phi_1^{}\;,\\
&&\\
\ds\phi_f & \equiv &
\ds\d_1'\*\phi_1'\;+\;\d_3'\*\phi_3' & = & \d_2^{}\*\phi_2^{}\;,\\
&&\\
\ds\phi_g & \equiv &
\ds\a_1'\*\phi_1'\;+\;\c_2'\*\phi_2' & = & \c_3^{}\*\phi_3^{}\;,
\end{array}\right.\end{equation}
\begin{equation}\label{phi-a}
\ds \phi_a \;\equiv\;\b_1'\*\phi_1'\;+\;\d_2'\*\phi_2'\;+\;
\c_3'\*\phi_3'\;=\;0\;.
\end{equation}
Given are the currents and the dynamical variables for the
left hand side graph. Due to $\phi_h=0$, eq. (\ref{phi-h}),
only two currents
are independent, let them be $\phi_1$ and $\phi_3$. 
All the variables
for the right hand side graph we try to restore via 
the linear system.
First, use $\phi_b$, $\phi_c$ and $\phi_d$ (\ref{phi-bcd})
to express all $\phi_j'$.
Substitute $\phi_j'$
into relations for $\phi_e$, $\phi_f$ and $\phi_g$ 
(\ref{phi-efg}), then it will appear
three homogeneous linear relations for 
two arbitrary $\phi_1$ and $\phi_3$,
so six coefficients of $\phi_1$ and 
$\phi_3$ must vanish. Solving this
six equations with respect to the 
primed variables, we obtain
\begin{equation}
\ds\left.
\begin{array}{cc}
\ds\b_2'\;\a_2^{\prime-1}\;=\;
\Lambda_1^{-1}\*\b_3^{}\;\a_3^{-1}\;,&
\ds\a_3'\;\b_3^{\prime-1}\;=\;
\Lambda_1^{-1}\*\a_2^{}\;\b_2^{-1}\;,\\
&\\
\ds\d_1'\;\c_1^{\prime-1}\;=\;
\Lambda_2^{-1}\*\b_3^{}\;\d_3^{-1}\;,&
\ds\d_3'\;\b_3^{\prime-1}\;=\;
\Lambda_2^{-1}\*\c_1^{}\;\d_1^{-1}\;,\\
&\\
\ds\a_1'\;\c_1^{\prime-1}\;=\;
\Lambda_3^{-1}\*\a_2^{}\;\c_2^{-1}\;,&
\ds\c_2'\;\a_2^{\prime-1}\;=\;
\Lambda_3^{-1}\*\c_1^{}\;\a_1^{-1}\;,
\end{array}\right.
\label{raz}
\end{equation}
where three polynomials arisen:
\begin{equation}\label{lambda-general}
\ds\left.\begin{array}{ccc}
\ds \Lambda_1^{} & = & 
\ds\b_3^{}\;\a_3^{-1}\;\a_1^{}\;\b_1^{-1}\;-\;
\c_1^{}\;\b_1^{-1}\;+\;\a_2^{}\;\b_2^{-1}\;
\d_1^{}\;\b_1^{-1}\;,\\
&&\\
\ds \Lambda_2^{} & = & 
\ds\b_3^{}\;\d_3^{-1}\;\c_2^{}\;\d_2^{-1}\;-\;
\a_2^{}\;\d_2^{-1}\;+\;\c_1^{}\;
\d_1^{-1}\;\b_2^{}\;\d_2^{-1}\;,\\
&&\\
\ds \Lambda_3^{} & = & 
\ds\a_2^{}\;\c_2^{-1}\;\d_3^{}\;\c_3^{-1}\;-\;
\b_3^{}\;\c_3^{-1}\;+\;\c_1^{}\;
\a_1^{-1}\;\a_3^{}\;\c_3^{-1}\;.
\end{array}\right.\end{equation}
Substituting $\phi_j'$ into $\phi_a=0$ (\ref{phi-a}),
we obtain the homogeneous linear
equation for $\phi_1$, $\phi_3$ again, 
and the coefficients of them
vanish if
\begin{equation}\label{dva}
\ds\left.\begin{array}{ccc}
\ds\b_1'\;\c_1^{\prime-1} & = & \ds\Lambda_a\;\Lambda_1\;\bigl(
\c_2^{}\;\b_2^{-1}\;\d_1^{}\;\b_1^{-1}\;+\;
\d_3^{}\;\a_3^{-1}\;\a_1^{}\;\b_1^{-1}\bigr)^{-1}\;,\\
&&\\
\ds\d_2'\;\a_2^{\prime-1} & = & \ds\Lambda_a\;\Lambda_2\;\bigl(
\a_1^{}\;\d_1^{-1}\;\b_2^{}\;\d_2^{-1}\;+\;
\a_3^{}\;\d_3^{-1}\;\c_2^{}\;\d_2^{-1}\bigr)^{-1}\;,\\
&&\\
\ds\c_3'\;\b_3^{\prime-1} & = & \ds\Lambda_a\;\Lambda_3\;\bigl(
\d_1^{}\;\a_1^{-1}\;\a_3^{}\;\c_3^{-1}\;+\;
\b_2^{}\;\c_2^{-1}\;\d_3^{}\;\c_3^{-1}\bigr)^{-1}\;,
\end{array}\right.\end{equation}
where $\Lambda_a$ is arbitrary. The origin of $\Lambda_a$
technically is $\phi_a\;=\;\Lambda_a\*\phi_h$.

This $\Lambda_a$ is a sort of a gauge. The origin of it is that
due to $\phi_a\;\equiv\;0$ we may change it $\phi_a\mapsto\lambda_a\phi_a$,
this gives $\Lambda_a\mapsto\lambda_a\Lambda_a$, or equivalent
\begin{equation}\label{lambda-a}
\ds
\b_1'\mapsto\lambda_a\,\b_1'\;,\;\;\;
\d_2'\mapsto\lambda_a\,\d_2'\;,\;\;\;
\c_3'\mapsto\lambda_a\,\c_3'\;.
\end{equation}
Analogous degree of freedom is lost in the map $W_1,W_2,W_3$ $\mapsto$
$W_1',W_2',W_3'$:
the system of the  observables is not changed when
$\phi_h\mapsto\lambda_h\phi_h$, i.e.  when
\begin{equation}\label{lambda-h}
\ds
\c_1\mapsto\lambda_h\,\c_1\;,\;\;\;
\a_2\mapsto\lambda_h\,\a_2\;,\;\;\;
\b_3\mapsto\lambda_h\,\b_3\;,
\end{equation}
and the formulae for $W_j'$ do not change with
(\ref{lambda-h}). Call such type invariance
of the system of the observables 
{\bf the site projective invariance}
(correspondingly, the site ambiguity of the dynamical variables).

The other obvious invariance (ambiguity) is 
{\bf the vertex projective}
one. As the consequence
of simple re-scaling of the currents almost nothing changes if
\begin{equation}\label{projective}
\ds
\a\mapsto\a\,\lambda\;,\;\;\;
\b\mapsto\b\,\lambda\;,\;\;\;
\c\mapsto\c\,\lambda\;,\;\;\;
\d\mapsto\d\,\lambda\;
\end{equation}
partially in all vertices $W_j$ and $W_j'$ with six 
different $\lambda_j$ and $\lambda_j'$.

Thus in the most general interpretation: 
the map $W_1,W_2,W_3$ $\mapsto$
$W_1',W_2',W_3'$
is defined up to projective ambiguity
$\lambda_1,\lambda_2,\lambda_3,\lambda_h$ $\mapsto$
$\lambda_1',\lambda_2',\lambda_3',\lambda_a$.

Very important feature of all these calculations is that

\begin{center}\fbox{we never tried
to commute anything !}\end{center}

Return to a general case of graph $G_n$.
$4\,N_V\;=\;2\,n\,(n-1)$
free invertible variables $\a_V,\b_V,\c_V,\d_V$, 
assigned to the vertices $V$ of
$G_n$, we regard as the generators of a body ${\cal B}(G_n)$.
Let ${\cal B}_P(G_n)$ be the set of functions on ${\cal B}(G_n)$
invariant with respect to the vertex ambiguity (\ref{projective}).
Note in general, for an open graph $G_n$ one may consider 
${\cal B}'_P(G_n)$ --
set of functions invariant with respect to both vertex 
and closed
site ambiguities. But such general considerations of 
${\cal B}'_P$
for the closed graphs, i.e. the graphs defined on the torus,
needs a notion of a trace (or of a characteristic polynomials),
or equivalent, of an algebra. The algebra  will be introduced 
in the subsequent section.

Consider a little change of $G_n$, so that only one $\lgr$ in
$G_n$ transforms into $\rgr$. Call the resulting graph $G_n'$. Let
the vertices involved into this change are marked as $W_1,W_2,W_3$
for $\lgr$ and $W_1',W_2',W_3'$ for $\rgr$ arranged as in Fig.
\ref{fig-YBE}. Introduce a functional operator $\R\;=\;\R_{1,2,3}$
making the corresponding map on ${\cal B}_P$:
\begin{equation}\label{themap}
\ds
\R_{1,2,3}^{}\cdot\varphi(W_1,W_2,W_3,...)\cdot
\R_{1,2,3}^{-1}\;=\;
\varphi(W_1',W_2',W_3',...)\;,\;\;\;\;
\varphi\in{\cal B}_P\;,
\end{equation}
where $W_j$ stands for $\{\a_j,\b_j,\c_j,\d_j\}$ forever,
and all other vertices except $W_1,W_2,W_3$ and their variables
remain untouched. This $\R$ we call the {\bf fundamental map}.

Let now $G_n'$ be an arbitrary graph equivalent to $G_n$. $G_n'$
can be obtained from $G_n$ by different sequences of elementary
$\lgr\mapsto\rgr$ in general. Thus the corresponding different
sequences of $\R$-s must coincide, this is natural admissibility
(or associability) condition for $G_n\mapsto G_n'$.

Note that in terms of functional operators the sequence of
na\"{\i}ve geometrical transformations is antihomomorphic to the
sequence of corresponding functional maps.

The simplest case is the equivalence of two quadrilaterals, $G_4$,
and the admissibility condition is nothing but the Tetrahedron
equation (\ref{te}). And due to the ambiguity of $\R$,
(\ref{lambda-a},\ref{lambda-h}), any admissibility condition is
still equation for $\Lambda_a$-th involved. Note, ${\cal
B}_P'(G_n)$ introduced previously, is gauge
invariant subspace of ${\cal B}_P$. $\R$ acts on ${\cal B}_P'$
uniquely. Unfortunately the basis of ${\cal B}_P'$ is not local,
and it is simpler to introduce an algebra constraints removing the
projective ambiguities then to consider ${\cal B}'_P$ formally.

{\bf A way to remove $\Lambda_a$ -- ambiguity from the definition
of $\R$, (\ref{raz},\ref{dva}), is to impose some additional
conditions for the elements of $W$, $\a,\b,\c,\d$, such that
(\ref{dva}) would become a consequence of (\ref{raz}) and the
additional conditions.}

Complete classification of these additional conditions is still
the open problem, and this is the main mathematical problem of
this approach.

\subsection{Local case: the Weyl algebra}

Here we consider a special {\em local} case: suppose first that
the elements of two different $W_i$ and $W_j$ for the given $G_n$
commute. Destroy also the vertex projective invariance
choosing $\a_j\equiv 1$
for any $j$ forever. Then (\ref{raz}) give the expressions for
$\b_2',\b_3'$, $\d_1'\c_1^{\prime-1},\d_3'\b_3^{\prime-1}$,
$\c_1',\c_2'$. Suppose also any pair of the variables from $W$
are linearly independent,
then
\begin{itemize}
\item
the commutability of the elements for different $W_j'$ from $\rgr$
gives (after some calculations) $\b\c=q\c\b$ with the same 
${\cal C}$-number $q$ for any vertex,
\item
these relations conserve by the map $\R$, 
i.e. $\b'\,\c'\;=\;q\;\c'\,\b'$.
\item
Also $\b^{-1}\c^{-1}\d$ appear to be centres, depending
on the vertex.
\end{itemize}
The gauge ambiguity becomes the ambiguity
for these centres.  We are going to get a sort of quantum theory,
$\b$ and $\c$ are already quantized, so we have to keep all 
centres to be invariant, 
$\b_j^{-1}\c_j^{-1}\d_j^{}\;=\;\b_j^{\prime-1}
\c_j^{\prime-1}\d_j^\prime$. This is possible, and
further we will threat these centres as a kind of spectral
parameters.

\begin{figure}
\begin{center}

\setlength{\unitlength}{0.25mm} 
\thicklines
\begin{picture}(450,200)
\put(125,0){
\begin{picture}(200,200)
\put(   0 ,   0 ){\vector( 1,1){200}}
\put( 200 ,   0 ){\vector(-1,1){200}}
\put( 100 , 100 ){\circle*{10}}
\put(  70 , 160 ){$q^{1/2}\;\;\u\*\phi$}
\put(  90 ,  30 ){$\w\*\phi$}
\put(  25 ,  95 ){$\phi$}
\put( 135 ,  95 ){$\kappa\;\;\u\*\w\*\phi$}
\end{picture}}
\end{picture}
\end{center}
\caption{Local parameterisation of the vertex}
\label{fig-weyl-vertex}
\end{figure}
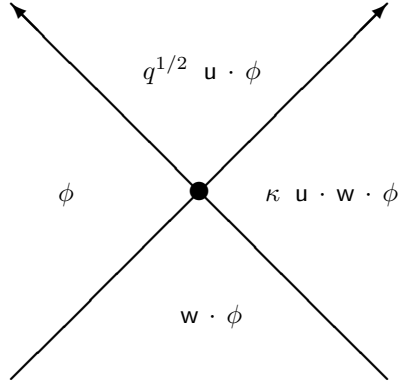

Change now notations for the dynamical variables to more
conventional, and write down the resulting expressions for the map
$\R$. New notations for the site currents are shown in
Fig.\ref{fig-weyl-vertex}.

\noindent
{\bf Proposition.} $\bullet$
Let the vertex dynamical variables are given by
\begin{equation}
\ds\a\;=\;1\;,\;\;\;\;
\b\;=\;q^{1/2}\;\u\;,\;\;\;
\c\;=\;\w\;,\;\;\;\;
\d\;=\;\kappa\;\u\;\w\;,
\end{equation}
here $\u,\w$ obey the local Weyl algebra relation,
\begin{equation}
\ds \u\*\w\;\;=\;\;q\;\;\w\*\u\;,
\end{equation}
and $\u$ and $\w$ for different vertices commute,
and number $\kappa$ is the invariant of the vertex, i.e.
$\kappa_{i,j}$, assigned to the intersection of lines $i$
and $j$, is the same for all equivalent graphs.

Then the problem of the algebraic equivalence
(i.e. equality of the outer currents)
of two graphs:
$G$ with the data $\phi,\u,\w$,
and $G'$ with the data $\phi',\u',\w'$,
can be solved
{\bf without any ambiguity} with respect to all $\phi',\u',\w'$,
and the local Weyl algebra structure for the set of
$\u',\w'$ is the consequence of the local Weyl algebra
relations for the set of $\u,\w$. $\bullet$

Write the fundamental simplex map for $\lgr=\rgr$ explicitly.
The map $\R=\R_{1,2,3}$ : $W_1,W_2,W_3$ $\mapsto$ $W_1',W_2',W_3'$,
\begin{equation}\label{R-action}
\R\*\u_j\;=\;\u_j'\*\R\;,\;\;\;
\R\*\w_j\;=\;\w_j'\*\R\;,\;\;\;j=1,2,3\;,
\end{equation}
is given by
\begin{equation}
\fbox{$\;\;\;
\ds\left.\begin{array}{ll}
&\\
\ds\w_1'\;=\;\ds\w_2^{}\* \Lambda_3^{}\;, &
\ds\u_1'\;=\;\ds\Lambda_2^{-1}\* \w_3^{-1}\;,\\
&\\
\ds\w_2'\;=\;\ds\Lambda_3^{-1}\* \w_1^{}\;, &
\ds\u_2'\;=\;\ds\Lambda_1^{-1}\* \u_3^{}\;,\\
&\\
\ds\w_3'\;=\;\ds\Lambda_2^{-1}\* \u_1^{-1}\;, &
\ds\u_3'\;=\;\ds\u_2^{}\* \Lambda_1^{}\;,\\
&
\end{array}\right.
\;\;\;$}
\label{ev}
\end{equation}
where
\begin{equation}\label{lambda}
\fbox{$\;\;\;
\ds\left.\begin{array}{ccl}
&&\\
\ds\Lambda_1 & = & \ds
\u_1^{-1}\*\u_3^{}\;-\;
q^{1/2}\;\u_1^{-1}\*\w_1^{}\;+\;
\kappa_1\;\;\w_1^{}\*\u_2^{-1}\;,\\
&&\\
\ds\Lambda_2 & = & \ds
{\kappa_1\over\kappa_2}\;\;\u_2^{-1}\*\w_3^{-1}\;+\;
{\kappa_3\over\kappa_2}\;\;\u_1^{-1}\*\w_2^{-1}\;-\;
q^{-1/2}\;\;{\kappa_1\;\kappa_3\over\kappa_2}\;\;
\u_2^{-1}\*\w_2^{-1}\;,\\
&&\\
\ds\Lambda_3 & = & \ds
\w_1^{}\*\w_3^{-1}\;-\;
q^{1/2}\;\;\u_3^{}\*\w_3^{-1}\;+\;
\kappa_3\;\;\w_2^{-1}\*\u_3\;.\\
&&
\end{array}\right.
\;\;\;$}
\end{equation}
Reverse formulae, giving $\R^{-1}$, look similar:
\begin{equation}
\ds
\left.\begin{array}{ccl}
\ds \Lambda_1^{-1} & = & \ds
{\kappa_1\over\kappa_2}\;\;\u_1^{\prime}\*\u_3^{\prime-1}\;-\;
q^{1/2}\;\;{\kappa_3\over\kappa_2}\;\;\u_1^{\prime}\*\w_1^{\prime-1}
\;+\;\kappa_3\;\;\w_1^{\prime-1}\*\u_2^{\prime}\;,\\
&&\\
\ds \Lambda_2^{-1} & = & \ds
\u_2^\prime\*\w_3^{\prime}\;+\;\u_1^\prime\*\w_2^\prime
\;-\;q^{-1/2}\;\;\kappa_2\;\;\u_2^\prime\*\w_2^\prime\;,\\
&&\\
\ds \Lambda_3^{-1} & = & \ds
{\kappa_3\over\kappa_2}\;\;\w_1^{\prime-1}\*\w_3^{\prime}
\;-\;q^{1/2}\;\;
{\kappa_1\over\kappa_2}\;\;\u_3^{\prime-1}\*\w_3^{\prime}
\;+\;\kappa_1\;\;\w_2^{\prime}\*\u_3^{\prime-1}\;.
\end{array}\right.
\end{equation}

The conservation of the Weyl algebra structure
\begin{equation}
\ds
\u_j\*\w_j\;=\;q\;\;\w_j\*\u_j\;\;\;\mapsto\;\;\;
\u_j'\*\w_j'\;=\;q\;\;\w_j'\*\u_j'
\end{equation}
means that $\R$ is the canonical map,
hence $\R_{1,2,3}$ can be regarded as an usual
operator depending on $\u_1,\w_1,\u_2,\w_2,\u_3,\w_3$.
The structure of $\R$
will be described in the next subsection.

Now the projective ambiguity is removed, and the current system
game gives the unique correspondence between the elements of
equivalent graphs. This is well defined meaning of the algebraic
equivalence. Hence all the admissibility conditions (and surely
the Tetrahedron relation) become trivial consequences of this
umambiguity, and we get them gratis !

Mention now a couple of useful limits of our fundamental map 
$\R_{1,2,3}$. The first one is the limit when 
$\kappa_1=\kappa_2=\kappa_3=\kappa$, and then $\kappa\mapsto 0$.
Denote such limiting procedure via
\begin{equation}\label{planar_conditions}
\ds
\kappa_1\;=\;\kappa_2\;=\;\kappa_3\;\;<<\;\;1\;.
\end{equation}
Corresponding map we denote $\R^{pl}_{1,2,3}$. The conditions
for $\kappa$-s are uniform for whole Tetrahedron relation,
\begin{equation}
\ds
\kappa_1\;=\;\kappa_2\;=\;\kappa_3\;=\;\kappa_4\;=\;\kappa_5\;=\;
\kappa_6\;\;<<\;\;1\;,
\end{equation}
so $\R^{pl}$ obeys TE. The other case is the limit of $\R_{1,2,3}$
when
\begin{equation}\label{one_conditions}
\ds
\kappa_1\;\;<<\;\;\kappa_2\;=\;\kappa_3\;\;<<\;\;1\;.
\end{equation}
These conditions are uniform for TE again,
\begin{equation}
\ds
\kappa_1\;\;<<\;\;\kappa_2\;=\;\kappa_3\;\;<<\;\;
\kappa_4\;=\;\kappa_5\;=\;\kappa_6\;\;<<\;\;1\;.
\end{equation}
Corresponding map we call $\mbox{\sf r}_{1,2,3}$,
and due to the uniformness it also obeys TE.
Recall, all these maps, $\R$ with 
$\kappa_1\;=\;\kappa_2\;=\;\kappa_3\;=\;1$, $\R^{pl}$
and $\mbox{\sf r}$ were derived previously as a hierarchy
of $\R$ -- operators solving TE, see \cite{sbm-qd,s-qd,ks-fun,
ms-modified}.

\subsection{Structure of $\R$}

Remarkable feature of $\R$ is its spatial invariance.
Change a little the operators on which $\R$ depends:
\begin{equation}\label{gammas}
\ds
\Gamma_1^{}\;=\;
\kappa_1^{-1}\;\;\u_2^{}\*\u_3^{-1}\*\Lambda_1^{}\;,\;\;\;\;
\Gamma_2^{}\;=\;
\kappa_2^{}\;\;\u_1^{}\*\w_3^{}\*\Lambda_2\;,\;\;\;\;
\Gamma_3^{}\;=\;
\kappa_3^{-1}\;\;\w_1^{-1}\*\w_2^{}\*\Lambda_3^{}\;.
\end{equation}
Then for $\ds\alpha,\beta,\gamma$ being the cyclic
permutations of $1,2,3$,
\begin{equation}
\ds\begin{array}{crc}
&\ds(\Gamma_\beta\cdot\Gamma_\alpha\;-\;q\;
\Gamma_\alpha\cdot\Gamma_\beta)\cdot
\Gamma_\gamma\;-\;\Gamma_\gamma\cdot
(\Gamma_\beta\cdot\Gamma_\alpha\;-\;q\;
\Gamma_\alpha\cdot\Gamma_\beta)& \\
&&\\
&\ds -\; q^{-1}\;(1-q)\;(1-q^2)\;
(\Gamma_\alpha\;-\;\Gamma_\beta) & =\;0\;,
\end{array}
\end{equation}
and
\begin{equation}
\ds\begin{array}{crc}
& \ds
q\;\Gamma_\alpha\cdot\Gamma_\beta\;-\;q^{-1}\;
\Gamma_\beta\cdot\Gamma_\alpha\;-\;
\Gamma_\gamma\;(q^{1/2}\;
\Gamma_\alpha\cdot\Gamma_\beta\;-\;q^{-1/2}\;
\Gamma_\beta\cdot\Gamma_\alpha) &\\
&&\\
& \ds +\;q^{-1/2}\;(1-q)\;(q^{-1}\;\Gamma_\alpha\;+\;q\;
\Gamma_\beta\;-\;\Gamma_\gamma) & =\;0\;.
\end{array}
\end{equation}
It resembles $SO(3)$ invariance.

Give now a realisation of $\R$ in terms of more 
simple functions. First, recall the definition 
and properties of the quantum dilogarithm.
Let conventionally
\begin{equation}
\ds (\x;q)_n\;=\;(1-\x)\;(1-q\x)\;(1-q^2\x)\;...\;(1-q^{n-1}\x)\;.
\end{equation}
Then the quantum dilogarithm (by definition) 
\cite{fk-qd,br-qd}
\begin{equation}
\ds
\psi(\x)\;\stackrel{def}{=}\; (q^{1/2}\x;q)_\infty\;=\;
\sum_{n=0}^\infty\;\; {(-1)^n\,q^{n^2/2}\over (q;q)_n}\;\x^n\;,
\end{equation}
and
\begin{equation}
\ds\psi(\x)^{-1}\;=\;\sum_{n=0}^\infty\;\;{q^{n/2}\over 
(q;q)_n}\;x^n\;.
\end{equation}
This function is useful for the rational transformations of 
the Weyl algebra:
\begin{equation}
\ds
\psi(q\x)\;=\;(1-q^{1/2}\x)^{-1}\;\psi(\x)\;,\;\;\;\;
\psi(q^{-1}\x)\;=\;(1-q^{-1/2}\x)\;\psi(\x)\;,
\end{equation}
hence
\begin{equation}
\ds
\psi(\u)\*\w\;=\;\w\*(1-q^{1/2}\u)^{-1}\*\psi(\u)\;,\;\;\;\;
\psi(\w)\*\u\;=\;\u\*(1-q^{-1/2}\w)\*\psi(\w)\;.
\end{equation}
$\psi$ is called the quantum dilogarithm due to the pentagon 
identity \cite{fk-qd}
\begin{equation}
\ds
\psi(\w)\*\psi(\u)\;=\;\psi(\u)\*\psi(-q^{-1/2}\;\u\;\w)\*\psi(\w)\;,
\end{equation}
this corresponds to Roger's five term relation for the usual 
dilogarithm.
From the other side $\psi$ is the quantum exponent due to
\begin{equation}
\ds \psi(\u)\*\psi(\w)\;=\;\psi(\u+\w)\;.
\end{equation}
Recall, everywhere the Weyl algebra relation 
$\u\,\w\;=\;q\,\w\,\u$ is implied.

Introduce now a generalised permutation function. Let
$\P(\x,\y)$, $\x\,\cdot\,\y\;=\;q^2\;\y\,\cdot\,\x$,
is defined by the following
relations:
\begin{equation}
\ds\left\{\begin{array}{ccccc}
\ds\P(q\;\x,\y) & = & \ds\y^{-1}\;\P(\x,\y) & = & \ds\P(\x,\y)\;\y\;,\\
&&&&\\
\ds\P(\x,q\;\y) & = & \ds\P(\x,\y)\;\x^{-1} & = & \ds\x\;\P(\x,\y)\;,
\end{array}\right.\end{equation}
and
\begin{equation}
\ds\P(\x,\y)^2\;=\;1\;.
\end{equation}
For $\z$ obeying
\begin{equation}
\ds
\x\*\z\;=\;q^{f_x}\;\z\*\x\;,\;\;\;\;
\y\*\z\;=\;q^{f_y}\;\z\*\y
\end{equation}
it follows
\begin{equation}
\ds
\P(\x,\y)\*\z\;=\;q^{f_x\,f_y}\;\;\z\*\x^{f_y}\*\y^{-f_x}\*\P(\x,\y)\;.
\end{equation}
This function we call the generalised permutation because of
usual permutation operator of the tensor product is
\begin{equation}
\ds
\P\;\equiv\;\P(\u\otimes\u^{-1},\w\otimes\w^{-1})\;\;.
\end{equation}

Considering three independent $\Gamma_\alpha$ (\ref{gammas}),
$\alpha=1,2,3$, one may see that
all them depends on three operators $\U$, $\W$ and $\s$:
\begin{equation}
\ds
\U\;=\;\w_{2}^{-1}\*\w_{3}^{}\;,\;\;\;\;
\W\;=\;\w_{1}^{}\*\u_{3}^{-1}\;,\;\;\;\;
-\;q^{1/2}\;\;\s\*\U\*\W^{-1}\;=\;\u_{1}^{}\*\u_{2}^{-1}\;.
\end{equation}
$\U\;\W\;=\;q\;\;\W\;\U$ and $\s$ is the center. 
One can directly verify that
\begin{equation}\label{R-op}
\ds
\fbox{$\ds
\R\;=\;\psi(\kappa_3\;\U)\*\psi(\W^{-1})\*
\P\bigl(\sqrt{\kappa_3\over\kappa_2}\;\;\U\;,\;\s^{-1}\*\W^{\,2}\bigr)\*
\psi({\kappa_1\over\kappa_3}\;\W)^{-1}\*
\psi(\kappa_2\;\U^{-1})^{-1}\;,
$}
\end{equation}
being substituted into (\ref{R-action}), 
gives (\ref{ev},\ref{lambda}).
On $\U$ and $\W$ , $\R$ acts as follows.
\begin{equation}
\ds\left\{\begin{array}{ccl}
\ds\R\*\U\*\R^{-1} & = & \ds
{\kappa_2\over\kappa_3}\;\;\U^{-1}\*
\bigl(\;\W\;-\;q^{-1/2}\;+\;\kappa_3\;\;\U\;\bigr)\*\\
&&\\
&&\ds\*
\bigl(\;\W\;-\;q^{-1/2}\;\;{\kappa_1\over\kappa_3}\;\;\s\;+\;
\kappa_1\;\;\;\s\*\U\;\bigr)^{-1}\;,\\
&&\\
\ds\R\*\W\*\R^{-1} & = & \ds
\s\*\W^{-1}\*
\bigl(\;\W\;-\;q^{1/2}\;+\;\kappa_3\;\;\U\;\bigr)\*
\bigl(\;\W\;-\;q^{1/2}\;+\;\kappa_1\;\;\s\*\U\;\bigr)^{-1}\;.
\end{array}\right.\end{equation}

When $\kappa_1=\kappa_2=\kappa_3=1$, expression (\ref{R-op})
for $\R$
coincides with the operator solution of the Tetrahedron equation
from \cite{s-qd,ms-modified}. This is the generalisation of the
finite dimensional 3D $R$-matrix from $q^N=1$ to general $q$, and
the finite dimensional $R$-matrix corresponds to the
Zamolodchikov--Bazhanov--Baxter model.

We don't discuss this correspondence here, the reader may find the
details concerning Zamolodchikov -- Bazhanov -- Baxter model in
\cite{zam-solution, baxter-pf,bb-first,kms-stsq}, the details
concerning the finite $R$-matrix -- in \cite{mss-vertex}, the
details concerning the quantum dilogarithm -- in original papers
\cite{fk-qd,br-qd}, and operator valued $\R$ as the generalisation
of finite $R$ -- in \cite{sbm-qd,s-qd,ms-modified,double}.


Few words concerning the meaning of (\ref{R-op}).
All $\psi$-s can be decomposed into the seria with respect to
their arguments. Substitute these $\R$-s into the Tetrahedron
relation (\ref{te}) and move all the generalized
permutations $\P$ out. $\P$-s theirself obey the 
Tetrahedron equation and so can be cancelled from TE for $\R$-s.
Then twelve $\psi$-s rest in the left hand side of TE, and twelve
$\psi$-s rest in the right hand side. The Tetrahedron equation
in this case becomes a relation resembling the braid group
relation in $2D$. This twenty-four terms relation can be 
proved {\bf directly} via the seria decomposition of all 
$24$ quantum dilogarithms. The proof is based on several
{\bf finite} $q$-re-summations (like $q$-binomial theorems).
This is the first value of the formula (\ref{R-op}).
The second one is that relation (\ref{R-op}) gives a 
nice way to derive the finite dimensional complete 
$R$-matrix (just replacing $\psi$-s and $\P$ by their
finite dimensional counterparts, \cite{br-qd,s-qd,ms-modified}).

Generalised permutation $\P(\x,\y)$ has no good series 
realization. Note, if we abolish condition $\P^2=1$ 
for a moment, then formally
\begin{equation}
\ds
\P(\x,\y)\;\sim\;\sum_{\alpha,\beta\in Z}\;\;
q^{-\alpha\,\beta}\;\x^\alpha\*\y^\beta\;.
\end{equation}
This $\P$ obeys $\P^2=1$ if one takes the Euler definition
$\ds\sum_{n\in Z}\;\;q^{n\;m}\;=\;\delta_{m,0}$ 
\cite{euler,struik}. Note that in the manipulations with $q$-seria
the Euler principle ``A sum of any infinite series is the value
of an expression, which expansion gives this series'' was never
failed.
Actually $\P(\x,\y)$ is to be defined specially for 
every realisation of the Weyl algebra. As an example
mention Kashaev and Faddeev's non invariant
realisation of the Weyl generators as shifts on the
space of appropriately defined functions $\varphi([t])$:
\begin{equation}
\ds
\w_j\*\varphi([t])\;=\;[t]_j\;\phi([t])\;,\;\;\;\;
\u_j\*\phi([t])\;=\;-\;q^{-1/2}\;[t]_j\;
\varphi([t]:[t]_j\,\mapsto\,q^{-1}\,[t]_j)\;.
\end{equation}
Where $[t]$ is a list of the arguments of $\varphi$ and
$[t]_j$ is its $j$-th component. Thus $\u_j$ and $\w_j$
refer to the $j$-th ``pointer'' of the list of arguments and
hence are not functional operators in usual sense.
Actually the action of $\u_j$, $\w_j$ on $\varphi([t])$
would be given symbolically by  the following correspondence:
\begin{equation}
\ds
\varphi([t])\;\;\leftrightarrow\;\;
|t_1>\otimes |t_2>\otimes |t_3>\otimes\;...\;,
\end{equation}
if  the eigenvectors $|t_j>$ of the operators $\w_j$
might be defined.

Generalised permutation introduced
\begin{equation}
\ds
\P_{1,2,3}\;=\;
P(\sqrt{\kappa_3\over\kappa_2}\;\U\;,\;\s^{-1}\,\W^2)\;=\;
\P(\sqrt{\kappa_3\over\kappa_2}\;\w_2^{-1}\,\w_3^{}\;,\;
-\,q^{-1/2}\,\u_1^{-1}\w_1^{}\,\w_2^{-1}\,\u_2^{}\,
\w_3^{}\u_3^{-1})
\end{equation}
act of $\u_j$, $\w_j$, $j=1,2,3$, as follows:
\begin{equation}
\ds\left\{
\begin{array}{ccl}
\ds\P_{1,2,3}\*\w_1 & = & \ds\sqrt{\kappa_2\over\kappa_3}\;
\w_1^{}\,\w_2^{}\,\w_3^{-1}\*\P_{1,2,3}\;,\\
&&\\
\ds\P_{1,2,3}\*\w_2 & = & \ds\sqrt{\kappa_3\over\kappa_2}\;
\w_3^{}\*\P_{1,2,3}\;,\\
&&\\
\ds\P_{1,2,3}\*\w_3 & = & \ds\sqrt{\kappa_2\over\kappa_3}\;
\w_2^{}\*\P_{1,2,3}\;,
\end{array}\right.
\end{equation}
and
\begin{equation}
\ds\left\{
\begin{array}{ccl}
\ds\P_{1,2,3}\*\u_1 & = & \ds\sqrt{\kappa_2\over\kappa_3}\;
\u_1^{}\,\w_2^{}\,\w_3^{-1}\*\P_{1,2,3}\;,\\
&&\\
\ds\P_{1,2,3}\*\u_2 & = & 
\ds -\;q^{1/2}\;\sqrt{\kappa_3\over\kappa_2}\;
\u_1^{}\,\w_1^{-1}\,\u_3^{}\*\P_{1,2,3}\;,\\
&&\\
\ds\P_{1,2,3}\*\u_3 & = & 
\ds-\;q^{1/2}\;\sqrt{\kappa_2\over\kappa_3}\;
\u_1^{-1}\,\w_1^{}\,\u_2^{}\*\P_{1,2,3}\;.
\end{array}\right.
\end{equation}
This gives the following action of $\P$ on 
$\varphi(t_1,t_2,t_3)$:
\begin{equation}
\ds
\P_{1,2,3}\*\varphi(t_1,t_2,t_3)\;=\;
\varphi(\sqrt{\kappa_2\over\kappa_3}\;{t_1\,t_2\over t_3}\;,\;
\sqrt{\kappa_3\over\kappa_2}\; t_3\;,\;
\sqrt{\kappa_2\over\kappa_3}\; t_2)\;,
\end{equation}
where $t_1,t_2,t_3$ stand on the positions corresponding
$1,2,3$ of $\R_{1,2,3}$.

Another thing to be mentioned is the case of $|q|=1$.
In this case the quantum dilogarithmic functions 
should be replaced by Faddeev's integral \cite{f-modular}.
In few words, it appears when one considers the Jacoby
imaginary transformation of an argument of $\psi$ and $q$:
\begin{equation}
\ds
\u\;=\;\mbox{\large e}^{i\,z}\;,\;\;\;\;
-\;q^{1/2}\;=\;\mbox{\large e}^{i\,\pi\,\theta}\;\;\;\mapsto\;\;\;
\widetilde{\u}\;=\;\mbox{\large e}^{i\,z/\theta}\;,\;\;\;\;
-\widetilde{q}^{1/2}\;=\;\mbox{\large e}^{-i\,\pi/\theta}\;.
\end{equation}
Then
\begin{equation}
\ds
\psi_{F}(\u)\;=\;{\ds(\;q^{1/2}\;\u\;;\;q\;)_\infty\over\ds
(\;\widetilde{q}^{1/2}\;\widetilde{\u}\;;
\;\widetilde{q}\;)_\infty}\;,
\end{equation}
and the following expression for $\psi_F(\u)$ is valid
in the limit of real $\theta$ \cite{f-modular}:
\begin{equation}
\ds
\psi_F(\u)\;(=\;s(z))\;=\;
\exp\;{1\over 4}\;\int_{\infty}^{\infty}
{\ds \mbox{\large e}^{z\,\xi}\over\ds\sinh \pi\xi\;
\sinh \pi\theta\xi}\;{\ds d\;\xi\over\ds \xi}\;,
\end{equation}
where the singularity at $\xi=0$ is circled from above.


\bigskip

Return now to map (\ref{R-op}).
The map $\R$ conserves four independent operators:
\begin{equation}
\ds
\w_1\cdot\w_2\;,\;\;\;\;\;\;
\u_2\cdot\u_3\;,\;\;\;\;\;\;\s
\end{equation}
and
\begin{equation}\label{H-op}
\ds\begin{array}{ccl}
\H & = &
\w_1^{}\*\u_3^{-1}
\;\;-\;\;
q^{1/2}\;\;\u_1^{}\*\u_2^{-1}\*\w_2^{}\*\w_3^{-1}\\
&&\\
& - &
\kappa_1\;\;q^{-1/2}\;\;\u_1^{}\*\w_1^{}\*\u_2^{-1}\*\u_3^{-1}
\;\;+\;\;
\kappa_3\;\;\u_1^{}\*\u_2^{-1}\\
&&\\
& - &
\kappa_2\;\;q^{-1/2}\;\;\w_1^{}\*\w_2^{}\*\u_3^{-1}\*\w_3^{-1}
\;\;+\;\;
\kappa_2\;\;\w_2^{}\*\w_3^{-1}\\
&&\\
& = & \ds
\biggl(\W^{-1}\;+\;\kappa_1\;\U\;-\;
q^{1/2}\;\kappa_3\;\U\;\W^{-1}\biggr)
\;+\;\s^{-1}\;
\biggl(\W\;+\;\kappa_2\;\U^{-1}\;-\;
q^{1/2}\;\kappa_2\;\U^{-1}\;\W\biggr)\;.
\end{array}
\end{equation}
Actually $\R$ depends only on two of them, $\s$ and $\H$.

Consider the following product
\begin{equation}\label{sigma-psi}
\ds\sigma\;=\;
\psi(a\;\w^{-1})\*\psi(b\;\u)\*
\psi(-\;q^{-1/2}\;c\;\u\;\w)\*
\psi(a'\;\w)\*\psi(b'\;\u^{-1})\;.
\end{equation}
Let
\begin{equation}
\ds
\chi\;=\;a\w^{-1}+a'\w+b\u+b'\u^{-1}-q^{-1/2}c\;\u\w
-q^{-1/2}ab'\;\u^{-1}\w^{-1}\;.
\end{equation}
It is easy to check $\sigma\*\chi\;=\;\chi\*\sigma$. Hence
$\sigma$ as an operator is a function on $\chi$:
\begin{equation}\label{sigma-chi}
\ds
\sigma\;=\;\sigma(\;aa'\;,\;bb'\;,\;{c\over a'b}\;\;|\;\chi\;)\;,
\end{equation}
I did not find explicit form of function $\sigma$,
only a special case of $\sigma$ when $c=b'=0$, then
\begin{equation}
\ds
\psi(a\,\w^{-1})\;\psi(b\,\u)\;\psi(a'\,\w)\;=\;
\psi(a\,\theta^{-1})\;\psi(a'\,\theta)
\end{equation}
where
\begin{equation}
\ds a\,\theta^{-1}\;+\;a'\,\theta\;=\;a\,\w^{-1}\;+\;b\,\u\;+\;a'\,\w\;.
\end{equation}
Nevertheless direct
calculations give $\R^2$ in terms of $\sigma$ introduced.
First, it is convenient to rewrite $\R$:
\begin{equation}
\ds
\R\;=\;\psi(\W^{-1})\,
\psi(-q^{1/2}\kappa_3\U\W^{-1})\,
\P(\sqrt{\kappa_3\over\kappa_2}\,\U,\s^{-1}\W^2)\,
\psi(-q^{1/2}{\kappa_1\kappa_2\over\kappa_3}\,\U^{-1}\W)^{-1}
\,\psi({\kappa_1\over\kappa_3}\,\W)^{-1}\;.
\end{equation}
Then
\begin{equation}
\ds
\R^2\;=\;\mbox{\cal N}\*\mbox{\cal D}^{-1}\;,
\end{equation}
where
\begin{equation}
\ds
\mbox{\cal N}\;=\;
\psi(\W^{-1})\,
\psi(-q^{1/2}\kappa_3\U\W^{-1})\,
\psi(\kappa_1\U)\,
\psi(\s^{-1}\W)\,
\psi(-q^{1/2}\kappa_2\s^{-1}\U^{-1}\W)\;,
\end{equation}
and
\begin{equation}
\ds
\mbox{\cal D}\;=\;
\psi({\kappa_1\over\kappa_3}\W)\,
\psi(-q^{1/2}{\kappa_1\kappa_2\over\kappa_3}\U^{-1}\W)\,
\psi({\kappa_1\kappa_2\over\kappa_3}\U^{-1})\,
\psi({\kappa_1\over\kappa_3}\s\W^{-1})\,
\psi(-q^{1/2}\kappa_1\s\U\W^{-1})\;.
\end{equation}
Comparing these with the definition of $\sigma$, we obtain
\begin{equation}
\ds
\mbox{\cal N}\;=\;
\sigma(\s^{-1},\kappa_2\kappa_3\s^{-1},
{\kappa_1\over\kappa_3}\s|\,\H\,)\;,\;\;\;\;
\mbox{\cal D}\;=\;
\sigma({\kappa_1^2\over\kappa_3^2}\s,
{\kappa_1^2\kappa_2\over\kappa_3}\s,
{\kappa_3\over\kappa_1}\s^{-1}|\,
{\kappa_1\over\kappa_3}\s\H\,)\;,
\end{equation}
where $\H$ is given by (\ref{H-op}).

\subsection{Fusion}

\begin{figure}
\begin{center}

\setlength{\unitlength}{0.25mm} 
\thicklines
\begin{picture}(450,200)
\put(110,0){
\begin{picture}(200,200)
\put(0,50){\vector(1,0){200}}\put(0,150){\vector(1,0){200}}
\put(50,0){\vector(0,1){200}}\put(150,0){\vector(0,1){200}}
\put(50,50){\circle*{5}}\put(50,150){\circle*{5}}
\put(150,50){\circle*{5}}\put(150,150){\circle*{5}}
\put(55,55){\scriptsize $1,2$}\put(55,155){\scriptsize $1,1$}
\put(155,55){\scriptsize $2,2$}\put(155,155){\scriptsize $2,1$}
\put(0,190){$\phi$}
\put(0,95){$y$}
\put(95,0){$x$}
\put(187,95){$-y'$}
\put(92,190){$-x'$}
\end{picture}}
\end{picture}
\end{center}
\caption{Combined vertex}
\label{fig-fusion}
\end{figure}
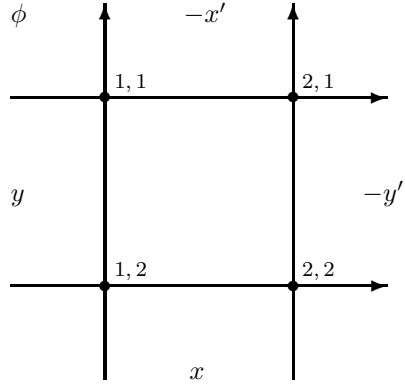

One more remarkable feature of the current model is a sort of 
a fusion. As an example consider a planar graph formed by two 
pairs of the parallel lines. Four vertices arise as the 
intersection points of these two pairs of the lines.
This is shown in Fig. \ref{fig-fusion}.

The vertices are labelled by the pairs of the indices,
$W_{1,1}$, $W_{1,2}$, $W_{2,1}$ and $W_{2,2}$. Single closed site
means that there are three independent currents. Let them be
the internal current $\phi_{1,1}$, assigned to north-west corner,
and two currents $x$ and $y$ assigned to  southern and western
semi-strips, $x$ and $y$ are observable currents for this cross
considered as an alone graph.

Applying the linear system rules, we obtain step by step
\begin{equation}
\ds\begin{array}{ccl}
\ds\phi_{1,1} & = & \ds\phi\;,\\
&&\\
\ds\phi_{1,2} & = & \ds y \;-\; \w_{1,1}^{}\*\phi\;,\\
&&\\
\ds\phi_{2,2} & = & \ds \w_{2,2}^{-1}\*x\;-\;
\kappa_{1,2}^{}\;\u_{1,2}^{}\;\w_{1,2}^{}\;\w_{2,2}^{-1}\*y\;+\;
\kappa_{1,2}^{}\;\w_{1,1}^{}\;\u_{1,2}^{}\;\w_{1,2}^{}\;\w_{2,2}^{-1}
\*\phi\;,\\
\end{array}\end{equation}
and from zero value of the closed site current
\begin{equation}
\ds\begin{array}{ccl}
\ds\phi_{2,1} & = & \ds
-\;\w_{2,1}^{-1}\;\w_{2,2}^{-1}\*x\;+\;
(\kappa_{1,2}^{}\;\u_{1,2}^{}\;\w_{1,2}^{}\;\w_{2,1}^{-1}\;\w_{2,2}^{-1}-
q^{1/2}\;\u_{1,2}^{}\;\w_{2,1}^{-1})\*y\\
&&\\
\ds & + & \ds
(q^{1/2}\;\w_{1,1}^{}\;\u_{1,2}^{}\;\w_{2,1}^{-1}-\kappa_{1,1}^{}\;
\u_{1,1}^{}\;\w_{1,1}^{}\;\w_{2,1}^{-1}-
\kappa_{1,2}^{}\;\w_{1,1}^{}\;\u_{1,2}^{}\;\w_{1,2}^{}\;
\w_{2,1}^{-1}\;\w_{2,2}^{-1})\*\phi\;.
\end{array}\end{equation}

\noindent
Let further $-x'$ and $-y'$ are the edge variables assigned
to the northern and eastern semistips. In general they are
\begin{equation}\label{matrixmap}
\ds \left\{\begin{array}{ccc}
\ds x' & = & \ds \alpha\* x\;+\;\beta\* y\;+\;\mbox{\sf f}_x\*\phi\;,\\
&&\\
\ds y' & = & \ds \gamma\* x\;+\;\delta\* y\;+\;\mbox{\sf f}_y\*\phi\;,
\end{array}\right.\end{equation}
where 
\begin{equation}
\ds\left\{\begin{array}{ccl}
\ds\alpha & = & \ds \w_{2,1}^{-1}\;\w_{2,2}^{-1}\;,\\
&&\\
\ds\beta & = & \ds q^{1/2}\;\u_{1,2}^{}\;\w_{2,1}^{-1}\;-\;
\kappa_{1,2}^{}\;\u_{1,2}^{}\w_{1,2}^{}\;\w_{2,1}^{-1}\;\w_{2,2}^{-1}\;,\\
&&\\
\ds\gamma & = & \ds -\;q^{1/2}\;\u_{2,2}^{}\;\w_{2,2}^{-1}\;+\;
\kappa_{2,1}^{}\;\u_{2,1}^{}\;\w_{2,2}^{-1}\;,\\
&&\\
\ds\delta & = & \ds q^{1/2}\;\kappa_{2,1}^{}\;\u_{1,2}^{}\;\u_{2,1}^{}
\;+\;q^{1/2}\;\kappa_{1,2}^{}\;
\u_{1,2}^{}\;\w_{1,2}^{}\;\u_{2,2}^{}\;\w_{2,2}^{-1}\\
&&\\
&&\ds -\;
\kappa_{1,2}^{}\;\kappa_{2,1}^{}\;
\u_{1,2}^{}\;\w_{1,2}^{}\;\u_{2,1}^{}\w_{2,2}^{-1}\;,
\end{array}\right.\end{equation}
and
\begin{equation}\label{feshki}
\ds\left\{\begin{array}{ccl}
\ds\mbox{\sf f}_x & = & \ds
-\;q^{1/2}\;\u_{1,1}^{}\;-\;
q^{1/2}\;\w_{1,1}^{}\;\u_{1,2}^{}\;\w_{2,1}^{-1}\\
&&\\
&& +\;\kappa_{1,1}^{}\;\u_{1,1}^{}\;\w_{1,1}^{}\;\w_{2,1}^{-1}
\;+\;\kappa_{1,2}^{}\;\w_{1,1}^{}\;\u_{1,2}^{}\;
\w_{1,2}^{}\;\w_{2,1}^{-1}\;\w_{2,2}^{-1}\;,\\
&&\\
\ds\mbox{\sf f}_y & = & \ds
-\;q^{1/2}\;\kappa_{2,1}^{}\;\w_{1,1}^{}\;\u_{1,2}^{}\;\u_{2,1}^{}\;-\;
q^{1/2}\;\kappa_{1,2}\;\w_{1,1}^{}\;\u_{1,2}^{}\;\w_{1,2}^{}\;
\u_{2,2}^{}\;\w_{2,2}^{-1}\\
&&\\
&&\ds+\;\kappa_{1,1}^{}\;\kappa_{2,1}^{}\;
\u_{1,1}^{}\;\w_{1,1}^{}\;\u_{2,1}^{}\;+\;
\kappa_{1,2}^{}\;\kappa_{2,1}^{}\;\w_{1,1}^{}\;\u_{1,2}^{}\;\w_{1,2}^{}\;
\u_{2,1}^{}\;\w_{2,2}^{-1}\;.
\end{array}\right.\end{equation}

Curents $x,y,x',y'$ become the edge currents when we rewrite
the cross in Fig. \ref{fig-fusion} 
as a single vertex with modified (thick)
lines; denote it as 
$\widetilde{W}\sim\{W_{1,1},W_{1,2},W_{2,1},W_{2,2}\}$.
Suppose we combine such crosses $\widetilde{W}$
(vertices with thick lines) in any way,
then zero value conditions for the restricted strips 
(closed thick edges)
look very simply: due to the signs $(-)$ in the definition of
outgoing
$x'$ and $y'$ these zero value conditions becomes
``outgoing edge current of one thick vertex $=$
incoming edge current of another thick vertex''.
Thus the strip variables just transfer from one
combined (thick) vertex to another, and therefore they
look like {\bf edge variables} of the thick vertices. 

In the case when for any thick vertex the map $x,y$ $\mapsto$
$x',y'$ (\ref{matrixmap}) does not contain extra $\phi$, i.e.
$\f_x=\f_y=0$ in any sense, then the part of the linear system
corresponding to the edge variables factorises from the whole
current system. 
{\bf If such factorisation exists for a graph $G$
then it exists for any equivalent graph $G'$, so sub-manifold of
${\cal B}_P$ given by $\f_x=\f_y=0$ is invariant of a map 
$G\mapsto G'$.}

On this sub-manifold we can delete all the edge currents 
$x=y=...=0$.
In this case all corner currents of cross $\widetilde{W}$
are proportional to
$\phi=\phi_{1,1}$, and the structure of the ``thick''
vertex becomes the
structure of usual vertex. Thus one may define ``thick'' analogies
of $\u,\w$ and $\kappa$. This phenomenon resembles the
usual two-dimensional fusion.

Write now explicit formulae. Introduce
\begin{equation}
\ds\begin{array}{ccl}
\ds\mbox{\sf K}^{-1} & = & \ds
{1\over\kappa_{1,2}^{}\kappa_{2,1}^{}\kappa_{2,2}^{}}\;\;\bigl(
\u_{1,1}^{}\;\u_{1,2}^{-1}\;+\;\w_{1,1}^{}\;\w_{2,1}^{-1}\;-\;
q^{-1/2}\;\;\kappa_{1,1}^{}\;
\u_{1,1}^{}\;\u_{1,2}^{-1}\;\w_{1,1}^{}\;\w_{2,1}^{-1}\bigr)\;,\\
&&\\
\ds\mbox{\sf k} & = & \ds q^{1/2}\;\;
\kappa_{2,1}^{}\;\kappa_{2,2}^{}\;
\w_{1,1}^{-1}\;\w_{1,2}^{-1}\;\w_{2,1}^{}\;\w_{2,2}^{}\;,\\
&&\\
\ds\widetilde{\mbox{\sf k}} & = & \ds q^{-1/2}\;\;
\kappa_{1,2}^{}\;\kappa_{2,2}^{}\;
\u_{1,1}^{-1}\;\u_{1,2}^{}\;\u_{2,1}^{-1}\;\u_{2,2}^{}\;.
\end{array}\end{equation}
Without mentioning of a representation of the Weyl algebra, its
right module etc., suppose $\phi$ in (\ref{matrixmap}) obeys
$\mbox{\sf f}_x\*\phi\;=\;\mbox{\sf f}_y\*\phi\;=\;0$.
From this, it follows that
\begin{equation}
\ds\mbox{\sf K}^{-1}\*\phi\;=\;\mbox{\sf k}^{-1}\*\phi\;=\;
\widetilde{\mbox{\sf k}}^{\;-1}\*\phi\;=\;K^{-1}\;\;\phi,
\end{equation}
where $K$ is introduced as an ``eigenvalue''.
On this ``subspace'' the fusion is defined as
\begin{equation}
\ds
\Delta(\w)\;=\;-\;\w_{1,1}^{}\;\w_{1,2}^{}\;,\;\;\;\;
\Delta(\u)\;=\;-\;q^{1/2}\;\u_{1,1}^{}\;\u_{2,1}^{}\;,\;\;\;\;
\Delta(\kappa)\;=\; K\;.
\end{equation}

The meaning of all these is the following. Consider
three ``thick'' crosses $\widetilde{W}_j\mapsto\widetilde{W}_j'$,
$j=1,2,3$,
arranged into ``thick'' Yang -- Baxter -- type
graphs, $\lgr$ and $\rgr$.
Solving the complete problem of the equivalence
(with twelve vertices in each hand side)
one obtains the set of relations like
\begin{equation}
\Delta(\w_j')\*\phi_j'\;-\;\Delta(\w_j)'\*\phi_j'\;=\;
\sum_{k=1}^3\;\;X_k\*\f_{x,k}\;\phi_k+Y_k\*\f_{y,k}\;\phi_k\;,
\end{equation}
etc., with some $X_k$ and $Y_k$, $\f_{x,k}$ and $\f_{y,k}$ given by
(\ref{feshki}).
$\Delta(\w_j')$ we obtain from $\Delta(\w_j)$ applying all
eight $\R$-s repeatedly, and $\Delta(\w_j)'$ is the result
of the application of single $\R$ in terms of $\Delta(\u_j)$,
$\Delta(\w_j)$ and $\Delta(K)$.

\subsection{Matrix part}

Few words concerning the matrix variables $\alpha,\beta,\gamma,\delta$ in
(\ref{matrixmap}). This remark is not important for our current
approach, but the structure of matrix variables 
is very interesting.
First, the map of edge auxiliary variables
\begin{equation}\label{korepanovmap}
\ds \left\{\begin{array}{ccc}
\ds x' & = & \ds \alpha\* x\;+\;\beta\* y\;,\\
&&\\
\ds y' & = & \ds \gamma\* x\;+\;\delta\* y\;,
\end{array}\right.\end{equation}
appeared in Korepanov's matrix models \cite{korepanov-diss,kks-fte}.
The Yang -- Baxter equivalence in Korepanov's interpretation
is the Korepanov equation: admissibility of the map of three
edge variables $(x,y,z)$ assigned to three lines of the Yang -- Baxter
graph. Let
\begin{equation}
\ds
X_1^{} \;=\;\left(\begin{array}{ccc}
\alpha_1 & \beta_1 & 0 \\ \gamma_1 & \delta_1 & 0 \\ 0 & 0 & 1
\end{array}\right)\;,\;\;\;\;
X_2^{} \;=\;\left(\begin{array}{ccc}
\alpha_2 & 0 & \beta_2 \\ 0 & 1 & 0 \\ \gamma_2 & 0 & \delta_2
\end{array}\right)\;,\;\;\;\;
X_3^{} \;=\;\left(\begin{array}{ccc}
1 & 0 & 0 \\ 0 & \alpha_3 & \beta_3 \\ 0 & \gamma_3 & \delta_3
\end{array}\right)\;,
\end{equation}
Then the admissibility is
\begin{equation}\label{KE}
\ds X_1^{}\* X_2^{}\* X_3^{}\;=\; X_3'\* X_2'\* X_1'\;,
\end{equation}
where the primed $X$-s 
consist on primed $\alpha,\beta,\gamma,\delta$. 
Korepanov's equation is
equivalent to the usual local Yang -- Baxter equation 
for the so-called
ferroelectric weights:
\begin{equation}
\ds X\;=\;\left(\begin{array}{cc} \alpha & \beta \\ 
\gamma & \delta\end{array}\right)
\;\;\;\;\mapsto\;\;\;\;
L\;=\;\left(\begin{array}{cccc}
1 & 0 & 0 & 0 \\
0 & \alpha & \beta & 0 \\
0 & \gamma & \delta & 0\\
0 & 0 & 0 & \zeta
\end{array}\right)\;,
\end{equation}
where in the numeric case $\zeta\;=\;\alpha\;\delta \;-\;\beta\;\gamma$, 
and the conventional $2^2\times 2^2$ $=$ $4\times 4$ 
matrix form for Yang -- Baxter matrix $L$
is used (for the equivalence see \cite{oneparam} for example).
In our case
the elements of different $X$-s commute, and the
elements of one $X$ obey the algebra
\begin{equation}\label{algebra}
\ds\left.\begin{array}{clc}
&\ds \alpha\*\beta\;=\;\beta\*\alpha\;,\;\;\;\; 
\ds\gamma\*\delta\;=\;\delta\*\gamma\;, &\\
&&\\
&\ds\alpha\*\gamma\;=\;q\;\;\gamma\*\alpha\;,\;\;\;\;
\ds\alpha\*\delta\;=\;q\;\;\delta\*\alpha\;,
\;\;\;\;
\ds\beta\*\delta\;=\;q\;\;\delta\*\beta\;,
\end{array}\right.\end{equation}
and
\begin{equation}\label{z-eq}
\ds
\zeta\;\stackrel{def}{=}\;
\alpha\*\delta\;-\;\beta\*\gamma\;=\;
\delta\*\alpha\;-\;\gamma\*\beta\;.
\end{equation}
From (\ref{algebra}) and (\ref{z-eq}) it follows that
$\zeta\;\beta=\beta\;\zeta$, $\zeta\;\gamma=\gamma\;\zeta$,
and consequently
\begin{equation}
\ds\left.\begin{array}{clc}
&\ds\beta^2\*\gamma\;+\;q\;\;\gamma\*\beta^2\;-\;
(1+q)\;\;\beta\*\gamma\*\beta\;=\;0\;,&\\
&&\\
&\ds\beta\*\gamma^2\;+\;q\;\;\gamma^2\*\beta\;-\;
(1+q)\;\;\gamma\*\beta\*\gamma\;=\;0\;.
\end{array}\right.\end{equation}
Hence
\begin{equation}\label{ad-bc}
\ds\alpha\*\delta\;=\;-\;{q\over 1-q}\;\;
(\beta\*\gamma\;-\;\gamma\*\beta)\;,
\end{equation}
and
\begin{equation}\label{z-bc}
\ds\zeta\;=\;-\;{1\over 1-q}\;\;
(\beta\*\gamma\;-\;q\;\;\gamma\*\beta)\;.
\end{equation}
Call the algebra of $\alpha,\beta,\gamma,\delta$, given by (\ref{algebra}) 
and (\ref{z-eq}), as ${\cal X}$.
Interesting is the following

\noindent
{\bf Proposition.} $\bullet$
Korepanov's equations are nine equation for twelve
variables, so $X_j'$ are defined ambiguously: in general one can't
fix one element from $\alpha_1^\prime,\alpha_2^\prime$, one from 
$\delta_2^\prime,\delta_3^\prime$, and one from 
$\alpha_3^\prime,\delta_2^\prime$. 
Impose on these three arbitrary elements the simple part of 
${\cal X}$, (\ref{algebra}).
Then all other relations of ${\cal X}$, namely relations
(\ref{algebra}) for the other primed elements and all
three relations (\ref{z-eq}) (or, equivalent, (\ref{ad-bc})) 
for the elements of $\{X_1',X_2',X_3'\}$, 
hold automatically as the consequence of
Korepanov's equation.
$\bullet$

This observation, we guess, 
is a way of a quantization of Korepanov's matrix model.
Remarkably is that ${\cal X}$ is the nontrivial algebra.

As an example consider the case when $\delta\;=\;0$.
Corresponding algebra, ${\cal X}_{\delta=0}$, contains only one nontrivial
relation, $\alpha\*\gamma\;=\;q\;\;\gamma\*\alpha$, 
and $\beta$ is a center.
Being a ${\cal C}$ -- number, $\beta_j$ are to be conserved by the
map $X_j^{}\mapsto X_j^\prime$. Equations (\ref{KE}) contain
$\beta_2\;=\;\beta_1\;\beta_3$. 
Hence $\beta$ is the pure gauge and one may put
$\beta\equiv 1$. The solution of (\ref{KE}) is:
\begin{equation}
\ds\left.\begin{array}{clc}
\ds & \left\{\begin{array}{ccl}
\ds\alpha_1^\prime & = & \ds
(\alpha_3^{}\;+\;\alpha_1^{}\*\gamma_3^{})^{-1}
\*\alpha_1^{}\*\alpha_2^{}\\
&&\\
\ds\gamma_1^\prime & = & \ds
f\*\gamma_1^{}\*\alpha_3^{}\*
(\alpha_3^{}\;+\;\alpha_1^{}\*\gamma_3^{})^{-1}
\end{array}\right. & \\
&\\
\ds & \left\{\begin{array}{ccl}
\ds\alpha_2^\prime & = & \ds
\alpha_3^{}\;+\;\alpha_1^{}\*\gamma_3^{}\\
&&\\
\ds\gamma_2^\prime & = & \ds\gamma_1^{}\*\gamma_3^{}
\end{array}\right. & \\
&\\
\ds & \left\{\begin{array}{ccl}
\ds\alpha_3^\prime & = & \alpha_2^{}\* f^{-1}\\
&&\\
\ds\gamma_3^\prime & = & 
(\alpha_3^{}\;+\;\alpha_1^{}\*\gamma_3^{})
\*\gamma_1^{-1}\*\gamma_2^{}\*\alpha_3^{-1}\* f^{-1}
\end{array}\right. &
\end{array}\right.
\end{equation}
where $f$ is not fixed by (\ref{KE}), this is the ambiguity mentioned.
Permutation relations for the combinations of the primed elements,
which do not contain $f$, namely for 
$\alpha_1^\prime$, $\alpha_2^\prime$,
$\gamma_2^\prime$, $\alpha_3^\prime\;\gamma_1^\prime$ and 
$\gamma_3^\prime\;\gamma_1^\prime$, do not contradict the set of
the local Weyl algebrae 
$\alpha_j^\prime\*\gamma_j^\prime\;=\;q\;\;
\gamma_j^\prime\*\alpha_j^\prime$. 
This corresponds to the statement of the proposition above. 
Consider now the Weyl algebrae for all primed elements. 
From this, it follows immediately
\begin{equation}
f\*\alpha_j\;=\;\alpha_j\* f\;,\;\;\;\;
f\*\gamma_j\;=\;\gamma_j\* f\;.
\end{equation}
Hence $f$ is a ${\cal C}$ -- number, and therefore we may put 
$f\;=\;1$. Thus the conservation of ${\cal X}_{\delta=0}$ fixes
the ambiguity.

The map $\alpha_j,\gamma_j \mapsto \alpha_j^\prime,\gamma_j^\prime$ 
we've obtained is nothing but $\mbox{\sf r}_{1,2,3}$, given by
the limiting procedure (\ref{one_conditions}). The identification is
$\alpha\;=\;\w^{-1}$ and $\gamma\;=\;-\;q^{1/2}\;\u^{}\cdot\w^{-1}$.
This case,
\begin{equation}
\ds X\;=\;\left(\begin{array}{rcr}
\w^{-1} &,& 1 \\ -\;q^{1/2}\;\u^{}\cdot\w^{-1} &,& 0
\end{array}\right)\end{equation}
is the quantization of the case $(\eta)$ from the list of simple
functional maps in \cite{oneparam}.

The case of general ${\cal X}$ is rather complicated technically,
it is a subject of a separate investigation.

\subsection{Co-current system and $\L$-operator}

In this subsection we give another form of the current approach.

Consider the whole linear system for a graph $G$ defined on a torus
(boundary conditions assumed). This system is the set of
zero equations
\begin{equation}\label{site-eq}
\ds
\phi_{\mbox{site}}\;\;\stackrel{def}{=}\;\;
\sum_{\mbox{vertices}}\;\;
W_{\mbox{vertex}}\*\phi_{\mbox{vertex}}\;\;=\;\;0\;,
\end{equation}
where such equation we write for each site of $G$,
the sum is taken over all vertices surrounded this site,
and contribution from vertex $V$, denoted as $W_V\cdot\phi_V$, 
is one of
$\phi_V$, $q^{1/2}\u_V\cdot\phi_V$, $\w_V\cdot\phi_V$  or
$\kappa_V\u_V\w_V\cdot\phi_V$ according to 
Fig. \ref{fig-weyl-vertex}
and the orientation of $V$.
The toroidal structure means that all the sites are 
closed and the number
of the sites equals to the number of the vertices.
Gathering all zero equations (\ref{site-eq}) together,
we obtain the matrix form of them,
\begin{equation}\label{LP}
\ds{\bf L}\*\Phi\;=\;0\;,
\end{equation}
where we combine the internal currents $\phi_V$
into the column $\Phi$ and the
matrix of the coefficients $\bf L$ consists of 
\begin{equation}
\ds
1\;,\;\;\; q^{1/2}\;\u_V\;,\;\;\; \w_V\;\;\;\mbox{and}\;\;\;
\kappa_V\;\u_V\;\w_V
\end{equation}
for all vertices $V$ of the lattice.
${\bf L}$ is the square matrix, and explicit form of it depends
on the geometry $G$.

(\ref{LP})
can be interpreted as an equation of motion for the action
${\cal A}\;=\;\Phi^*\*{\bf L}\*\Phi$, 
where the row co-current vector
$\Phi^*$ does not depend on $\Phi$ and its components $\phi^*_S$
are assigned to the sites $S$ of $G$. The equation of motion for
$\Phi^*$ is $\Phi^*\*{\bf L}\;=\;0$.
Corresponding zero equations now are assigned to the vertices,
and each such equation connects the site co-currents from the 
sites surrounding this vertex.
The problem of the equivalence of
$\lgr$ and $\rgr$ in terms of co-currents can be 
formulated as follows:
the co-current system for the left hand side graph $\lgr$ of Fig.
\ref{fig-YBE} is
\begin{equation}\label{co-lhs}
\ds\left\{\begin{array}{ccccc}
\ds\phi^*_1 & \equiv & \ds\phi^*_c\;+
\;\phi^*_e\*q^{1/2}\;\u_1^{}\;+\;
\phi^*_h\*\w_1^{}\;+
\;\phi^*_d\*\kappa_1^{}\;\u_1^{}\;\w_1^{} & = & 0\;,\\
&&&&\\
\ds\phi^*_2 & \equiv & \ds\phi^*_h\;+
\;\phi^*_d\*q^{1/2}\;\u_2^{}\;+\;
\phi^*_b\*\w_2^{}\;+
\;\phi^*_f\*\kappa_2^{}\;\u_2^{}\;\w_2^{} & = & 0\;,\\
&&&&\\
\ds\phi^*_3 & \equiv & \ds\phi^*_c\;+
\;\phi^*_h\*q^{1/2}\;\u_3^{}\;+\;
\phi^*_g\*\w_3^{}\;+
\;\phi^*_b\*\kappa_3^{}\;\u_3^{}\;\w_3^{} & = & 0\;,
\end{array}\right.\end{equation}
and co-current system for the right hand side graph $\rgr$ is
\begin{equation}\label{co-rhs}
\ds\left\{\begin{array}{ccccc}
\ds\phi^{*\prime}_1 & \equiv & \ds
\phi^*_g\;+\;\phi^*_a\*q^{1/2}\;\u_1'\;+\;
\phi^*_b\*\w_1'\;+\;\phi^*_f\*\kappa_1^{}\;\u_1'\;\w_1' 
& = & 0\;,\\ &&&&\\
\ds\phi^{*\prime}_2 & \equiv & \ds
\phi^*_c\;+\;\phi^*_e\*q^{1/2}\;\u_2'\;+\;
\phi^*_g\*\w_2'\;+\;\phi^*_a\*\kappa_2^{}\;\u_2'\;\w_2' 
& = & 0\;,\\ &&&&\\
\ds\phi^{*\prime}_3 & \equiv & \ds
\phi^*_e\;+\;\phi^*_d\*q^{1/2}\;\u_3'\;+\;
\phi^*_a\*\w_3'\;+\;\phi^*_f\*\kappa_3^{}\;\u_3'\;\w_3' & = & 0\;.
\end{array}\right.\end{equation}
The equivalence means that when we remove 
$\phi^*_h$ from (\ref{co-lhs})
and $\phi^*_a$ from (\ref{co-rhs}), 
then the resulting systems as the
systems for $\phi^*_b,...,\phi^*_f$ are equivalent.

\begin{figure}
\begin{center}
\setlength{\unitlength}{0.25mm} 
\thicklines
\begin{picture}(200,200)
\put(  50 , 50 ){\vector(1,1){100}}
\put( 150 , 50 ){\vector(-1,1){100}}
\put( 100 , 100 ){\circle*{10}}
\put(  30 ,  20 ){$\y$}
\put(  30 , 160 ){$\x'$}
\put( 160 ,  20 ){$\x$}
\put( 160 , 160 ){$\y'$}
\put( 110 , 95 ){$\kappa,\u,\w$}
\end{picture}
\end{center}
\caption{Edge variables.}
\label{fig-edge}
\end{figure}
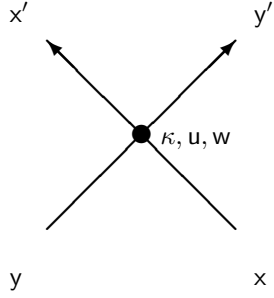

Consider now co-current equation for single vertex, as in Figs.
\ref{fig-weyl-vertex} or \ref{fig-abcd-currents}. 
Let the co-currents be $\phi^*_a$, $\phi^*_b$, 
$\phi^*_c$ and $\phi^*_d$, where the indices
$a,b,c,d$ are arranged as in Fig. \ref{fig-abcd-currents}. 
The co-current equation for this vertex is
\begin{equation}\label{co-eq}
\ds\phi^*\;\equiv\;\phi^*_a\;+\;\phi^*_b\* q^{1/2}\;\u\;+\;
\phi^*_c\*\w\;+\;\phi^*_d\*\kappa\;\u\;\w\;\;=\;\;0\;.
\end{equation}
Suppose we have solved a part of such equations for 
whole graph $G$,
and obtain $\phi^*_a$, $\phi^*_c$, $\phi^*_d$ 
in the form usual for
homogeneous linear equations:
\begin{equation}
\ds
\phi^*_a\;=\;-\;\phi^*_c\* q^{1/2}\;\y\;,\;\;\;\;
\phi^*_c\;=\;-\;\phi^*_d\* q^{1/2}\;\x
\end{equation}
with some multipliers $\x$ and $\y$. Then from (\ref{co-eq}) 
we get
\begin{equation}
\ds\phi^*_b\;=\;-\;\phi^*_d\* q^{1/2}\;\y'\;,\;\;\;
\mbox{or}\;\;\;\phi^*_a\;=\;-\;\phi^*_b\* q^{1/2}\;\x'\;,
\end{equation}
where
\begin{equation}
\ds\x'\;=\;\omega^{-1}\*\y\;,\;\;\mbox{and}\;\;\y'\;=\;\x\*\omega
\end{equation}
with
\begin{equation}\label{omega}
\ds\omega\;=\;\omega(\,\x\,,\,\y\,|\,\u\,,\,\w\,)\;=\;
\y\*\u^{-1}\;-\;q^{1/2}\;\;\u^{-1}\*\w\;+\;\kappa\;\;\x^{-1}\*\w
\;.
\end{equation}
Now we may change the interpretation completely. 
Assign $\x$, $\y$,
$\x'$, $\y'$ to the edges which separates corresponding sites.
These edge variables are shown in Fig. \ref{fig-edge}.

Now we can introduce the auxiliary functional operator $\L$,
giving the map $\x,\y$ $\mapsto$ $\x',\y'$, as we used to be:
\begin{equation}\label{L-op}
\ds\fbox{$\ds\;\;\;
\left.\begin{array}{ccc}
&&\\
\ds \L_{\x,\y}(\kappa,\u,\w)\*\x & = &
\omega(\,\x\,,\,\y\,|\,\u\,,\,\w\,)^{-1}\*
\y\*\L_{\x,\y}(\kappa,\u,\w)\;,\\
&&\\
\ds \L_{\x,\y}(\kappa,\u,\w)\*\y & = &
\;\;\x\*
\omega(\,\x\,,\,\y\,|\,\u\,,\,\w\,)\*
\;\L_{\x,\y}(\kappa,\u,\w)\;.\\
&&
\end{array}\right.
\;\;\;$}
\end{equation}
With the definition (\ref{R-action}), $\L$ operators obey
\begin{equation}\label{RLLL}
\ds\begin{array}{ccc}
& \ds
\L_{\y,\z}(\kappa_3,\u_3,\w_3)\*\L_{\x,\z}(\kappa_2,\u_2,\w_2)\*
\L_{\x,\y}(\kappa_1,\u_1,\w_1)\*\R_{1,2,3}\;= &\\
&&\\
&\ds =\;
\R_{1,2,3}\*\L_{\x,\y}(\kappa_1,\u_1,\w_1)\*
\L_{\x,\z}(\kappa_2\u_2,\w_2)\*
\L_{\y,\z}(\kappa_3,\u_3,\w_3) & \;.
\end{array}
\end{equation}
Moreover, Local Yang-Baxter relation
\begin{equation}\label{LLL}
\ds\begin{array}{ccc}
&\ds
\L_{\y,\z}(\kappa_3,\u_3,\w_3)\*\L_{\x,\z}(\kappa_2,\u_2,\w_2)\*
\L_{\x,\y}(\kappa_1,\u_1,\w_1)\;=&\\
&&\\
&\ds =\;
\L_{\x,\y}(\kappa_1,\u_1',\w_1')\*
\L_{\x,\z}(\kappa_2,\u_2',\w_2')\*
\L_{\y,\z}(\kappa_3,\u_3',\w_3')
\end{array}
\end{equation}
as a set of relations for $\u_k',\w_k'$, with $\u_k,\w_k$ given and
with $\x,\y,\z$ arbitrary,
gives again the map (\ref{ev}) uniquely ! Thus the kind of
the local Yang -- Baxter relation appears and for our 
current approach.

Conclude this section by few remarks 
concerning the functional maps.
All the maps introduced are connected 
to several graphical manipulations.
Usually we combine such manipulations 
($\lgr$ $\mapsto$ $\rgr$ of $\x,\y$
$\mapsto$ $\x',\y'$ etc.), and write 
the sequence of the
dynamical variables' sets obtained 
$\Sigma$ $\mapsto$ $\Sigma'$, in the
direct form
\begin{equation}
\ds
\Sigma\;=\;\Sigma_0\;
\stackrel{\ds A_1}{\mapsto}\;\Sigma_1\;
\stackrel{\ds A_2}{\mapsto}\;\Sigma_2\;\dots\;\Sigma_{n-1}\;
\stackrel{\ds A_n}{\mapsto}\;\Sigma_n\;,
\end{equation}
where $A_j$ stands for $j$-th manipulation, 
which allows us to calculate
$\Sigma_j$ in terms of previous variables 
$\Sigma_{j-1}$. The same
result, $\Sigma_0\mapsto\Sigma_n$, can be 
obtained as
\begin{equation}
\ds
\A_1\;\A_2\;...\;\A_n\*\Sigma_0\;=
\;\Sigma_n\*\A_1\;\A_2\;...\;\A_n\;,
\end{equation}
where $\A_j$ is a functional operator corresponding 
the manipulation $A_j$. 
Remarkable is the reverse order of the operators with respect to
the na\"{\i}ve manipulations.
Note that the direct order we obtain
considering the ``pointer'' action of the operators,
as it was mentioned in the previous subsection,
but the ``pointer'' action is not suitable for the 
quantization.


\section{Evolution system}

In this section we apply operator $\R$ defined in the previous
section to construct an evolution model explicitly.
Due to the current system's
background we formulate this model in terms of the regular
lattice defined on the torus, its motion, its current system and so on.

The main result of our paper is the generating function for the integrals
of motion for the evolution. The derivation of the integrals is based on
the auxiliary linear problem.

\subsection{Kagome lattice on the torus}

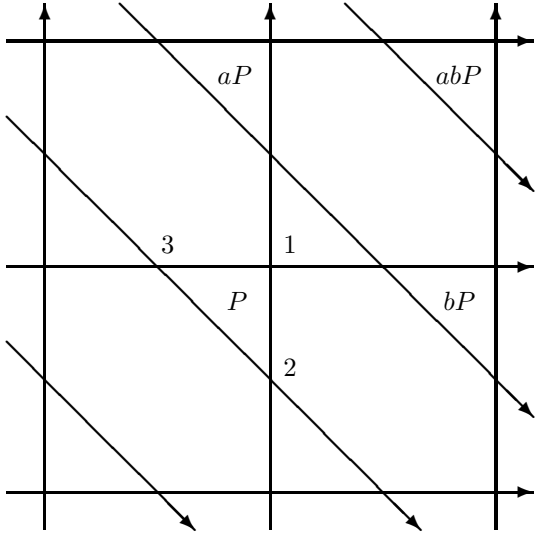
\begin{figure}
\begin{center}
\setlength{\unitlength}{0.25mm} 
\thicklines
\begin{picture}(450,280)
\put(85,0){
\begin{picture}(280,280)
\multiput(20,0)(120,0){3}{\vector(0,1){280}}
\multiput(0,20)(0,120){3}{\vector(1,0){280}}
\put(0,100){\vector(1,-1){100}}
\put(0,220){\vector(1,-1){220}}
\put(60,280){\vector(1,-1){220}}
\put(180,280){\vector(1,-1){100}}
\put(147,147){$1$}
\put(82,147){$3$}
\put(147,82){$2$}
\put(117,117){$P$}
\put(112,237){$aP$}
\put(232,117){$bP$}
\put(228,237){$abP$}
\end{picture}}
\end{picture}
\end{center}
\caption{The kagome lattice.}
\label{fig-kagome}
\end{figure}

An example of a regular lattice which contains both $\lgr$ and
$\rgr$ -- type triangles is so-called kagome lattice.
As it was mentioned in the introduction, the kagome lattices
appear in the sections of the regular 3D cubic lattices by 
inclined planes.
Thus the kagome lattice and its evolution corresponds actually
to the rectangular 3D lattice and thus is quite natural.
The kagome lattice
consists on three sets of parallel lines,  
usual  situation shown in Fig.
\ref{fig-kagome}. 
The sites of the lattice are both $\lgr$ and $\rgr$
triangles, and hexagons.

For given lattice introduce the labelling
for the vertices.
Mark the $\lgr$ triangles by the point notation $P$,
and let $a$ and $b$ are the multiplicative shifts in the northern
and eastern directions, so that the elementary shift in the 
south-east direction is $c=a^{-1}b$.
Nearest to triangle $P$ are
triangles $aP$, $bP$, $cP$, $a^{-1}P$, $b^{-1}P$
and $c^{-1}P$. 
Some of them are shown in Fig \ref{fig-kagome}.

For three vertices surrounding the $\lgr$-type triangle $P$
introduce the notations $(1,P)$, $(2,P)$ and $(3,P)$. 
These notations we will use as the subscripts for 
everything assigned to the vertices.

This kagome lattice we define on the torus of size $M$,
formally this means
the following equivalence:
\begin{equation}
\ds a^M\;P\;\sim\;b^M\;P\sim\;c^M\;P\;\sim\;P\;.
\end{equation}

\begin{figure}
\begin{center}

\setlength{\unitlength}{0.25mm} 
\thicklines
\begin{picture}(450,200)
\put(00,0){
\begin{picture}(200,200)
\put(0,150){\vector(1,0){200}}\put(50,150){\circle*{5}}
\put(0,200){\vector(1,-1){200}}\put(150,50){\circle*{5}}
\put(150,0){\vector(0,1){200}}\put(150,150){\circle*{5}}
\put(160,160){$1,P$}
\put(55,160){$3,P$}
\put(160,55){$2,P$}
\put(105,110){$\u_j,\w_j$}
\end{picture}}
\put(220,80){$\ds\stackrel{\U}{\mapsto}$}
\put(250,0){
\begin{picture}(200,200)
\put(50,0){\vector(0,1){200}}\put(50,50){\circle*{5}}
\put(0,50){\vector(1,0){200}}\put(50,150){\circle*{5}}
\put(0,200){\vector(1,-1){200}}\put(150,50){\circle*{5}}
\put(20,30){$1,P$}
\put(60,155){$2,a\,P$}
\put(155,60){$3,b\,P$}
\put(60,75){$\u_j',\w_j'$}
\end{picture}}
\end{picture}
\end{center}
\caption{Geometrical representation of evolution.}
\label{fig-U}
\end{figure}
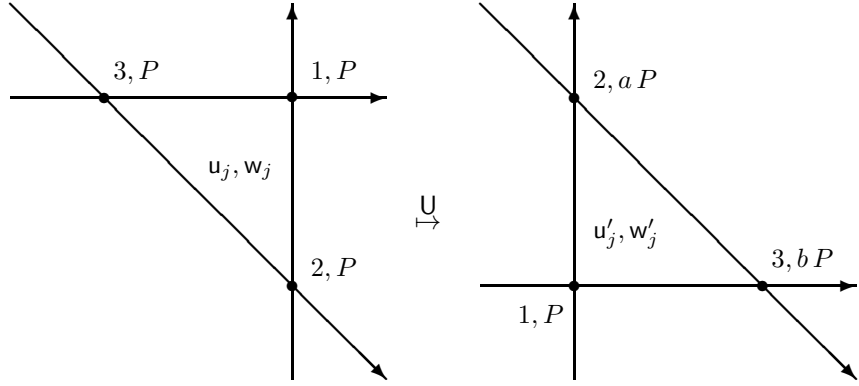

Since the notion of the equivalence, 
we may consider the shifts af all
inclined lines through the rectangular 
vertices into north-eastern direction 
as it is shown in Fig. \ref{fig-U}. 
It is easy to see that Fig. \ref{fig-U} 
is equivalent to Fig. \ref{fig-YBE}. 
The structure of the kagome lattice 
conserves by such shifts being made 
simultaneously for all $\lgr$-s, 
but the marking of the vertices changes a little.
This is visible in Fig. \ref{fig-U}.

Give now pure algebraic definition of the evolution. 
The phase space of the system is the set of $3\;M^2$ 
Weyl pairs $\u_{j,P}$ and $\w_{j,P}$,
$j=1,2,3$, $P = a^\alpha\,b^\beta\,P_0$, 
where $P_0$ is some frame of the reference's 
distinguished point, and the toroidal boundary conditions
mean
\begin{equation}
\ds\left.\begin{array}{ccccc}
\ds\u_{j,a^M\,P} & = & \ds\u_{j,b^M\,P} & = & \ds\u_{j,P}\;,\\
&&&&\\
\ds\w_{j,a^M\,P} & = & \ds\w_{j,b^M\,P} & = & \ds\w_{j,P}\;.
\end{array}\right.\end{equation}
The phase space is quantized by the definition.
Let $\u_{j,P}',\w_{j,P}'$ for $P$ fixed are given 
by (\ref{ev}), so that the map $\{\u_{j,P},\w_{j,P}\}$ 
$\mapsto$ $\{\u_{j,P}',\w_{j,P}'\}$
is given by the operator
\begin{equation}
\ds
\ds{\cal R}\;=\;\prod_{P}\;\;\R_P\;,
\end{equation}
where $\R_{P'}$ acts trivially on the variables of any 
triangle $P\neq P'$. 
Note, we suppose $\kappa_{j,P}$ 
do not depend on $P$, 
\begin{equation}
\ds\kappa_{j,P}\;=\;\kappa_{j}\;,
\end{equation}
so that with respect to $\kappa$-s
the translation invariance of the 
lattice is assumed. Define the superscript
`$\smb$' as follows:
\begin{equation}\label{smb}
\ds \left\{\begin{array}{ll}
\ds\u_{1,P}^\smb\;=\;\u_{1,P}'\;,&\ds\w_{1,P}^\smb
\;=\;\u_{1,P}'\;,\\ &\\
\ds\u_{2,aP}^\smb\;=\;\u_{2,P}'\;,&\ds\w_{2,aP}^\smb
\;=\;\w_{2,P}'\;,\\ &\\
\ds\u_{3,bP}^\smb\;=\;\u_{3,P}'\;,&\ds\w_{3,bP}^\smb
\;=\;\w_{3,P}'\;.
\end{array}\right.\end{equation}
This identification means following: 
$\u_{j,P}^\smb,\w_{j,P}^\smb$
are the variables
which appear on the places of previous 
$\u_{j,P},\w_{j,P}$ according to
Fig. \ref{fig-U}. The evolution operator
$\U$ : $\{\u_{j,P},\w_{j,P}\}$ $\mapsto$ 
$\{\u_{j,P}^\smb,\w_{j,P}^\smb\}$
we define as usual:
\begin{equation}\label{u-star}
\ds
\U\*\u_{j,P}^{}\*\U^{-1}\;=\;\u_{j,P}^\smb\;,\;\;\;\;
\U\*\w_{j,P}^{}\*\U^{-1}\;=\;\w_{j,P}^\smb\;.
\end{equation}
Regard the primary variables $\{\u_{j,P},\w_{j,P}\}$ 
of the given lattice
as the initial data for the discrete time evolution,
\begin{equation}
\ds\u_{j,P}\;=\;\u_{j,P}(0)\;,\;\;\;\;
\w_{j,P}\;=\;\w_{j,P}(0)\;.
\end{equation}
The evolution from $t=n$ to $t=n+1$ is just
\begin{equation}\label{u-nn}
\ds\u_{j,P}(n+1)\;=\;\U\*\u_{j,P}(n)\*\U^{-1}\;,\;\;\;\;
\w_{j,P}(n+1)\;=\;\U\*\w_{j,P}(n)\*\U^{-1}\;.
\end{equation}
Surely, the map $\U$ is the canonical map for the Weyl algebrae,
so that $\U$ is the quantum evolution operator.
Further we'll consider mainly the situation for $t=0$ 
and the map from $t=0$ to $t=1$. We will omit the time variable 
and write $f$ instead of $f(0)$ and $f^\smb=\U\* f\* U^{-1}$ 
instead of $f(1)$ for any object $f$.  Due to the homogeneity of
evolution (\ref{u-nn},\ref{u-star},\ref{smb}) our considerations 
appear to be valid for a situation with $t=n$ and the map
from $t=n$ to $t=n+1$.

\subsection{Linear system}

Investigate now the linear system for the quantum system obtained.

Assign to the vertex $(j,P)$ of the primary ($t=0$) 
kagome lattice
the internal current $\phi_{j,P}$. The linear system is the
set of $3\,M^2$ linear
homogeneous equation for $3\,M^2$ internal currents
\begin{equation}\label{f-values}
\ds\left\{\begin{array}{ccccc}
\ds f_{1,P} & \equiv & \ds
\w_{1,P}\*\phi_{1,P}+\phi_{2,P}+q^{1/2}\u_{3,P}\*\phi_{3,P}
& = & 0\;,\\
&&&&\\
\ds f_{2,P} & \equiv & \ds
q^{1/2}\u_{1,P}\*\phi_{1,P}+
\kappa_2\u_{2,aP}\w_{2,aP}\*\phi_{2,aP}
+\w_{3,bP}\*\phi_{3,bP} & = & 0\;,\\
&&&&\\
\ds f_{3,P} & \equiv & \ds
\phi_{1,a^{-1}P}+
\kappa_1\u_{1,b^{-1}P}\w_{1,b^{-1},P}\*\phi_{1,b^{-1}P}
+\w_{2,P}\*\phi_{2,P}&&\\
&&&&\\
&&\ds +q^{1/2}\u_{b,b^{-1}P}\*\phi_{2,b^{-1}P}
+\phi_{3,a^{-1}P}+\kappa_3\u_{3,P}\w_{3,P}\*\phi_{3,P} & = & 0\;.
\end{array}\right.
\end{equation}
Here we have introduced absolutely unessential notations $f_{j,P}$
just in order to distinguish these equations. $f_{j,P}$ are 
assigned to the sites.
Due to the homogeneity we may impose 
{\bf the quasiperiodical boundary
conditions} for $\phi_{j,P}$:
\begin{equation}\label{quasiperiod}
\ds
\phi_{j,a^M\,P}\;=\;A\;\phi_{j,P}\;,\;\;\;\;
\phi_{j,b^M\,P}\;=\;B\;\phi_{j,P}\;.
\end{equation}
It is useful to rewrite this system in the 
matrix form as (\ref{LP}),
$F\;\equiv\;{\bf L}\*\Phi\;=\;0$.
First combine $\phi_{j,P}$ with the same $j$  
into the column vector
$\Phi_j$ with $M^2$ components, 
so as $(\Phi_j)_P\;=\;\phi_{j,P}$.
Introduce matrices $T_a$ and $T_b$
as
\begin{equation}
\ds
(T_a\*\Phi_j)_P\;=\;\phi_{j,a\,P}\;,\;\;\;
(T_b\*\Phi_j)_P\;=\;\phi_{j,b\,P}\;.
\end{equation}
Due to (\ref{quasiperiod})
\begin{equation}
\ds T_a^M\;=\;A\;,\;\;\; T_b^M\;=\;B\;.
\end{equation}
Combine further $\u_{j,P}$ and $\w_{j,P}$ with the same $j$ into
diagonal matrices $\u_j$ and $\w_j$ with the same ordering 
of $P$ as in the definition of $\Phi_j$, 
\begin{equation}
\ds
\u_{j}\;=\;\mbox{diag}_P\;\; u_{j,P}\;\;,\;\;\;\;
\w_{j}\;=\;\mbox{diag}_P\;\; w_{j,P}\;\;.
\end{equation}
Obviously,
\begin{equation}
\ds
(T_a^{}\*\u_j\* T_a^{-1})_P\;=\;\u_{j,a\,P}\;,\;\;\;\;
(T_b^{}\*\u_j\* T_b^{-1})_P\;=\;\u_{j,b\,P}\;,
\end{equation}
and the same for $\w_j$.

Combine further $\Phi_1$, $\Phi_1$, $\Phi_3$ into $3\,M^2$ column $\Phi$.
Then from (\ref{f-values}) the matrix ${\bf L}$ can be extracted in the
$3\times 3$ $M^2\times M^2$ block form:
\begin{equation}\label{L-matrix}
\ds\fbox{
$\ds
{\bf L}\;=\;
\left(\begin{array}{rcrcr}
\w_1 &,& 1 &,& q^{1/2}\;\u_3\\
&&&&\\
q^{1/2}\;\u_1 &,& T_a\;\kappa_2\;\u_2\;\w_2 &,& T_b\;\w_3 \\
&&&&\\
T_a^{-1}\+ T_b^{-1}\;\kappa_1\;\u_1\;\w_1 &,&
\w_2\+T_b^{-1}\;q^{1/2}\;\u_2 &,&
T_a^{-1}\+\kappa_3\;\u_3\;\w_3
\end{array}\right)
$
}
\end{equation}
Recall, system ${\bf L}\*\Phi\;=\;0$ is $3\,M^2$ equations for
$3\,M^2$ components of $\Phi$.

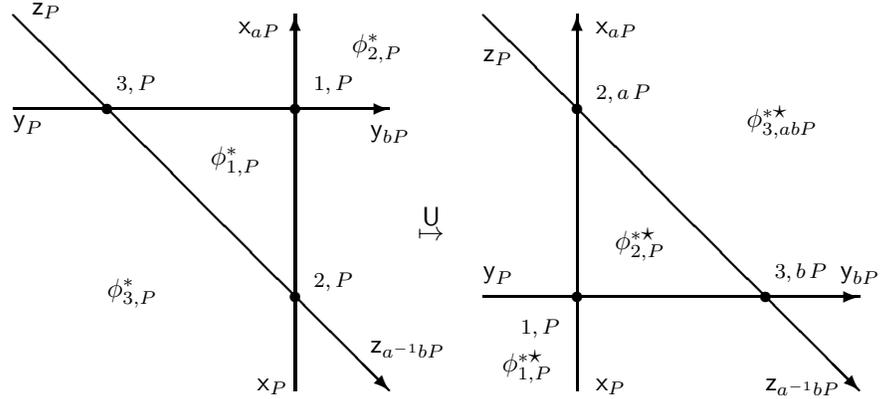
\begin{figure}
\begin{center}
\setlength{\unitlength}{0.25mm} 
\thicklines
\begin{picture}(450,200)
\put(00,0){
\begin{picture}(200,200)
\put(0,150){\vector(1,0){200}}\put(50,150){\circle*{5}}
\put(0,200){\vector(1,-1){200}}\put(150,50){\circle*{5}}
\put(150,0){\vector(0,1){200}}\put(150,150){\circle*{5}}
\put(160,160){\scriptsize $1,P$}
\put(55,160){\scriptsize $3,P$}
\put(160,55){\scriptsize $2,P$}
\put(105,120){$\phi^*_{1,P}$}
\put(180,180){$\phi^*_{2,P}$}
\put(50,50){$\phi^*_{3,P}$}
\put(10,200){$\z_P$}
\put(0,140){$\y_P$}
\put(130,0){$\x_P$}
\put(190,20){$\z_{a^{-1}bP}$}
\put(190,135){$\y_{bP}$}
\put(120,190){$\x_{aP}$}
\end{picture}}
\put(220,80){$\ds\stackrel{\U}{\mapsto}$}
\put(250,0){
\begin{picture}(200,200)
\put(50,0){\vector(0,1){200}}\put(50,50){\circle*{5}}
\put(0,50){\vector(1,0){200}}\put(50,150){\circle*{5}}
\put(0,200){\vector(1,-1){200}}\put(150,50){\circle*{5}}
\put(20,30){\scriptsize $1,P$}
\put(60,155){\scriptsize $2,a\,P$}
\put(155,60){\scriptsize $3,b\,P$}
\put(70,75){$\phi^{*\smb}_{2,P}$}
\put(10,10){$\phi^{*\smb}_{1,P}$}
\put(140,140){$\phi^{*\smb}_{3,abP}$}
\put(0,60){$\y_P$}
\put(0,175){$\z_P$}
\put(60,0){$\x_P$}
\put(190,60){$\y_{bP}$}
\put(150,0){$\z_{a^{-1}bP}$}
\put(60,190){$\x_{aP}$}
\end{picture}}
\end{picture}
\end{center}
\caption{Co-currents on the lattice}
\label{fig-cocurrents}
\end{figure}

Introduce now co-currents. As it was mentioned, 
${\bf L}\*\Phi\;=\;0$
we regard as the equations of motion for 2D 
system with the action
\begin{equation}
\ds {\cal A}\;\equiv\;\Phi^*\*{\bf L}\*\Phi\;.
\end{equation}
The block form of the co-currents $\Phi^*$ is 
thus fixed from the form of
${\bf L}$, or from (\ref{f-values}).
Equations of motion for $\Phi^*$ are 
$F^*\;\equiv\;\Phi^*\*{\bf L}\;=\;0$,
and in the component form
\begin{equation}\label{co-f-prime-values}
\ds\left\{\begin{array}{ccc}
\ds f^{*}_{1,P} & \equiv & \ds
\phi^{*}_{1,P}\*q^{1/2}\;\u_{1,P}^{}+
\phi^*_{2,b^{-1}P}+
\phi^*_{2,a^{-1}P}\*\kappa_1\;\u_{1,P}^{}\;\w_{1,P}^{}+
\phi^*_{3,P}\*\w_{1,P}^{}\;,\\
&&\\
\ds f^{*}_{2,P} & \equiv & \ds
\phi^{*}_{1,P}\*\kappa_2\;\u_{2,P}^{}\;\w_{2,P}^{}+
\phi^*_{2,P}\*q^{1/2}\;\u_{2,P}^{}+
\phi^*_{2,b^{-1}P}\*\w_{2,P}^{}+
\phi^*_{3,aP}\;,\\
&&\\
\ds f^{*}_{3,P} & \equiv & \ds
\phi^{*}_{1,P}\*\w_{3,P}^{}+
\phi^*_{2,P}+
\phi^*_{2,a^{-1}P}\*\kappa_3\;\u_{3,P}^{}\;\w_{3,P}^{}+
\phi^*_{3,bP}\*q^{1/2}\;u_{3,P}^{}\;.
\end{array}\right.
\end{equation}
Here $f_{j,P}^*$ corresponds to $(j,P)$-th vertex. The assignment
of the co-currents is shown in Fig. \ref{fig-cocurrents}.

Elements of $F^*\;=\;\Phi^*\*{\bf L}$ have the following 
remarkable
feature: coefficients in $f^*_{j,P}$ belong to the algebra of
$\u_{j,P}$, $\w_{j,P}$ only. We will use this in the next 
subsection.

\subsection{Properties of ${\bf L}$ and the quantum determinant}

Consider first the general properties of equation 
$\Phi^*\*{\bf V}\;=\;0$
for a matrix ${\bf V}$ similar to ${\bf L}$ introduced:
\begin{equation}
\ds {\bf V}\;=\;||\v_{j,k}||\;\;,
\end{equation}
with the commutative columns,
\begin{equation}\label{commutativity}
\ds \forall j,j'\;:\;\;\;\;
\v_{j,k}\* \v_{j',k'}\- \v_{j',k'}\* \v_{j,k}\;=\;0\;\;\;
\mbox{if}\;\;\; k'\neq k\;.
\end{equation}
Such matrices have the following properties.

\noindent
{\bf Property} 1: Consider a system
\begin{equation}
\ds \sum_{j}\;z_j\*\v_{j,k}\;=\;\alpha_k
\end{equation}
with $\alpha_k$ being $C$ -- numbers, as the system for $z_j$.
Then for $k\neq k'$
\begin{equation}
\ds
\alpha_k\;\alpha_{k'}\-\alpha_{k'}\;\alpha_k\;=\;
\sum_{j'}\; z_{j'}\;\alpha_k\;\v_{j',k'}\-
\sum_j\; z_j\;\alpha_{k'}\;\v_{j,k}\;=\;
\sum_{j,j'}\; (z_{j'}z_{j}-z_{j}z_{j'})\;\v_{j,k}\;\v_{j',k'}
\;=\;0\;.
\end{equation}
Matrix $||\v_{j,k}\*\v_{j',k'}||$ is non-degenerative in general,
so the last equality gives immediately
\begin{equation}\label{z-com}\ds
z_{j}\* z_{j'}\;=\; z_{j'}\* z_{j}\;.
\end{equation}

\noindent
{\bf Consequence}: Let $||\widetilde{\v}_{i,j}||$ 
is the inverse to
$||\v_{j,k}||$ matrix:
\begin{equation}
\ds \sum_{j}\;\widetilde{\v}_{i,j}\*  \v_{j,k}\;=\;
\sum_{j}\;\v_{i,j}\*\widetilde{\v}_{j,k}\;=\;\delta_{i,k}\;,
\end{equation}
then
\begin{equation}
\ds \forall i\;\;\;\;
\widetilde{\v}_{i,j}\*\widetilde{\v}_{i,j'}\;-\;
\widetilde{\v}_{i,j'}\*\widetilde{\v}_{i,j}\;=\;0\;.
\end{equation}

\noindent
{\bf Property} 2: Because of in $||\v_{j,k}||$ non-commutative 
elements
belong to the same column, the algebraic supplements $V_{k,l}$
as well as the quantum determinant $\mbox{\sf det}\;( \v)$ 
are well defined.
Here we've used the notation ``$\mbox{\sf det}$'' as the formal
operator-valued determinant
\begin{equation}
\ds \mbox{\sf det}\;||\v_{i,j}||\;=\;
\sum_{\sigma}\;(-1)^\sigma\;\prod_{j}\;\v_{j,\sigma(j)}\;.
\end{equation}
$V_{i,j}$ and $\mbox{\sf det}\;(\v)$ are polynomials of $\v_{j,k}$
such that in each summand
all multipliers belong to different columns and thus commute.
Moreover, if in $||\v||$ two rows coincide,
then $\mbox{\sf det}\;(\v)\;\equiv\;0$. Hence
\begin{equation}
\ds
\sum_{k}\;\; \v_{j,k}\* V_{k,l}\;=\;\delta_{j,l}\;\;
\mbox{\sf det}\;(\v)\;.
\end{equation}
Note, $\ds \v_{j,k}\* V_{k,l}\;=\;V_{k,l}\*\v_{j,k}$.

As it was mentioned
previously, sometimes it is useful to introduce formally a
module for the body of $||\v_{j,k}||$. Here we do this, 
introducing
$\phi^*_j$ and $\phi^*_0$ which belong to such formal module.
This allows us to formulate the following

\noindent
{\bf Consequence}:
Consider now the system of co-vector equations
\begin{equation}\label{veceq}
\ds
(\Phi^*\*{\bf V})_k\;=\;\sum_{j}\;\;\phi^*_j\*\v_{j,k}\;=\;0\;.
\end{equation}
Due to property 2 all $\phi^*_j$ belong to the null space of
$\mbox{\sf det}\;(\v)$:
\begin{equation}
\ds\phi^*_j\*\mbox{\sf det}\;(\v)\;=\;0\;.
\end{equation}
From the other hand side, $\phi^*_{j}$-s are connected by
$z_{j,j'}$ -- some rational functions of $\v_{j,k}$:
\begin{equation}
\ds
\phi^*_{j'}\;=\;\phi^*_{j}\* z_{j,j'}\;.
\end{equation}
Property 1 provides the commutability of $z_{j,j'}$, hence
a solution of (\ref{veceq}) can be written as
\begin{equation}\label{properties}
\ds
\phi^*_j\;=\;\phi^*_0\*z_j\;,\;\;\;
z_j\*z_{j'}\;=\;z_{j'}\*z_j\;,\;\;\;
\phi^*_0\*\mbox{\sf det}\;(\v)\;=\;0\;,\;\;\;
z_j\*\mbox{\sf det}\;(\v)\;=\;\mbox{\sf det}\;
(\v)\*\widetilde z_j\;,
\end{equation}
where in general $z_j\;\neq\;\widetilde z_j$.

Apply now both properties and their consequences to 
${\bf L}$ given by
(\ref{L-matrix}). First, for any representation of 
the Weyl algebrae
the null subspace $\phi^*_0$ of whole Gilbert space is defined,
\begin{equation}
\ds \phi^*_0\*\mbox{\sf det}\;({\bf L})\;=\;0\;.
\end{equation}
The existence of $\phi^*_0$ means the solvability of
$\Phi^*\*{\bf L}\;=\;0$. Corresponding $z_j$ have 
the lattice structure,
$z_{j,P}$. These commutative elements are assigned 
to the sites of the
kagome lattice, and observable are 
$z_{j,P}^{}\;z_{j'.P'}^{-1}$.
These operators connect the co-currents in 
different sites, and thus
$z_{j,P}$ actually give the realisation of 
the path group on the kagome lattice.

Another important thing is that due to 
$T_a^M\;=\;A$ and $T_b^M\;=\;B$,
$\mbox{\sf det}({\bf L})$
is a Laurent polynomial with respect to 
the quasimomenta
$A$ and $B$.

\subsection{Evolution of the co-currents and integrals of motion}

Consider now the shift of the inclined lines giving the evolution.
The internal currents as well as the co-currents change, 
and we can trace these changes.

Introduce two extra matrices, ${\bf K}$ and ${\bf M}$:
\begin{equation}\label{K-matrix}
\ds
{\bf K}\;=\;
\left(\begin{array}{rcrcr}
0 &,& \Lambda_0 &,& 0\\
&&&&\\
0 &,& 0 &,& T_a\;T_b \\
&&&&\\
1 &,& K_{3,2} &,& 0
\end{array}\right)\;,
\end{equation}
where
\begin{equation}
\ds
\Lambda_0\;=\;{\kappa_1\over\kappa_2}
\;q^{-1/2}\;\w_1^{}\,\u_2^{-1}\,\w_3^{-1}\;+\;
{\kappa_3\over\kappa_2}\;\u_1^{-1}\,\w_2^{-1}\,\u_3^{}\;,
\end{equation}
\begin{equation}
\ds
K_{3,2}\;=\;
T_a^{-1}\;\;q^{-1/2}\;\;\Lambda_2\+{\kappa_3\over\kappa_2}
\;\;\Lambda_1\+
T_b^{-1}\;\;{\kappa_1\over\kappa_2}\;\;\Lambda_3\;.
\end{equation}
with $\Lambda_j$ standing for the diagonal matrices with
the entries given by(\ref{ev}) correspondingly, and
\begin{equation}\label{M-matrix}
\ds
{\bf M}\;=\;
\left(\begin{array}{rcrcr}
0 &,& \u_1^{-1}\;\u_2'\;T_a &,& q^{-1/2}\;\u_1^{-1}\;T_b\\
&&&&\\
\ds{\kappa_1\over\kappa_2}\;\w_2^{-1}\;
\u_2^{-1}\;\u_1'\;\w_1' &,& 0 &,&
\ds{\kappa_3\over\kappa_2}\;\w_2^{-1}\;
\u_2^{-1}\;\u_3'\;\w_3'\;T_b\\
&&&&\\
\w_3^{-1} &,& \w_3^{-1}\;\w_2'\;T_a &,& 0
\end{array}\right)\;.
\end{equation}
\bigskip
Apply the evolution operator $\U$ to $\bf L$:
$\ds {\bf L}^\smb\;\equiv\;\U\*{\bf L}\*\U^{-1}$,
\begin{equation}
\ds
{\bf L}^\smb\;=\;
\left(\begin{array}{rcrcr}
\w_1' &,& 1 &,& q^{1/2}T_b^{-1}\u_3'T_b^{}\\
&&&&\\
q^{1/2}\u_1' &,& \kappa_2\u_2'\w_2'T_a^{} &,& \w_3'T_b^{} \\
&&&&\\
T_a^{-1}+ T_b^{-1}\kappa_1\u_1'\w_1' &,&
T_a^{-1}(\w_2'+T_b^{-1}q^{1/2}\u_2')T_a^{} &,&
T_a^{-1}+T_b^{-1}\kappa_3\u_3'\w_3'T_b^{}
\end{array}\right).
\end{equation}
\bigskip
\noindent
The following relation can be verified directly:
\begin{equation}\label{KLM}
\ds
{\bf K}\*{\bf L}^\smb\;=\;
{\bf L}\* {\bf M}\;.
\end{equation}
$\bf M$ in general is the matrix making
$\phi_{k,P}^\smb\mapsto\phi_{k,P}$, and $\bf K$ makes 
$\phi^*_{k,P}\mapsto
\phi^{*\smb}_{k,P}$.
Also ${\bf K}$ and ${\bf M}$ admit
\begin{equation}
\ds
{\bf K}\;\mapsto\;{\bf K}\+ {\bf L}\* {\bf N}\;,\;\;\;\;
{\bf M}\;\mapsto\;{\bf M}\+ {\bf N}\* {\bf L}^\smb
\end{equation}
with arbitrary ${\bf N}$. One can prove the following

\noindent
{\bf Proposition}: $\bullet$ ${\bf K}\*\mbox{\sf det}({\bf L})\;=\;
\mbox{\sf det}({\bf L})\*\widetilde{\bf K}$ $\bullet$

One can understand this in other terms. 
Since $\Phi^*\*{\bf L}\;=\;0$
can be solved for $t=0$, then for $t=1$ equation
$\Phi^{*\smb}\*{\bf L}^\smb\;=\;0$ must also be 
solved because they are
bounded by simple linear relations. Hence subspace $\phi^*_0$
must coincide with $(\phi^*_0)^\smb$, i.e.
\begin{equation}\label{dets}
\ds
\mbox{\sf det}({\bf L}^\smb)\;=\;\mbox{\sf det}({\bf L})\*D\;,
\end{equation}
with some operator $D$. One may hope, $D$ is not too complicated, 
and (\ref{dets}) is not trivial.

Careful analysis of ${\bf K}$ and ${\bf M}$ shows that 
this $D$ does not
depend on the quasimomenta $A$ and $B$. 
In the functional limit $q^{1/2}\mapsto\pm 1$ one may
easily calculate the determinants of ${\bf K}$ and ${\bf M}$,
both them are proportional to $A^M\;B^M$, and this term cancels
from the determinants of the left and right hand sides of 
(\ref{KLM}). This is so and in the quantum case.

Hence $D$ in (\ref{dets}) is a ratio of any
$A,B$ -- monomials from $\mbox{\sf det}({\bf L})$ and
$\mbox{\sf det}({\bf L}^\smb)$. Element $D$ can be extracted,
say, from $A^M\;B^{-M}$ component of $\mbox{det} ({\bf L})$:
\begin{equation}
\ds
D\;=\;\prod_{P}\;\;\u_{1,P}^{-1}\*
\prod_{P}\;\;\u_{1,P}^\smb\;.
\end{equation}
This means that we can introduce a simple
operator $d$:
\begin{equation}
\ds D\;=\;d\*d^{\smb -1}\;.
\end{equation}
Thus
\begin{equation}
\ds\J\;=\;\mbox{\sf det} \;(\;{\bf L}\;)\* d
\end{equation}
is the invariant of the evolution, $\J^\smb\;=\;\J$, i.e.
\begin{equation}
\ds\U\*\J\;=\;\J\*\U\;.
\end{equation}
Decompose $\J$ as a series of $A$ and $B$,
\begin{equation}
\ds\J\;=\;\sum_{\alpha,\beta\in\Pi}
\;\;A^\alpha\;B^\beta\;\J_{\alpha,\beta}\;,
\end{equation}
where $\alpha$ and $\beta$ are integers and their domain
(Newton's polygon) $\Pi$ is defined by
$|\alpha|\leq M$, $|\beta|\leq M$ and $|\alpha+\beta|\leq M$.
Quasimomenta $A$ and $B$ are arbitrary ${\cal C}$ -- numbers,
and the invariance of $\J$ means the invariance of each 
$\J_{\alpha,\beta}$. From the other side, 
$\J$ is a functional of the dynamical
variables of the lattice, i.e. 
\begin{equation}
\ds
\J_{\alpha,\beta}\;=\;\J_{\alpha,\beta}(
\{\u_{j,P}\,,\,\w_{j,P}\})\;.
\end{equation}
Surely, due to the homogeneity of the lattice
these functionals does not depend on time layer,
and hence the conservation of $\J$, $\J^\smb\;=\;\J$, means
\begin{equation}
\ds
\J_{\alpha,\beta}(\{\u_{j,P},\w_{j,P}\})\;=\;
\J_{\alpha,\beta}(\{\u_{j,P}^\smb,\w_{j,P}^\smb\})\;,
\end{equation}
i.e. functionals $\J_{\alpha,\beta}$ give the integrals of
motion in usual sense.

Note further, 
\begin{equation}\label{U-on-z}
\ds\Phi^*\*{\bf K}\;\sim\;\Phi^*\*\U^{-1}\;,
\end{equation}
where it is supposed $\phi^*_0\*\U^{-1}\;\sim\;\phi^*_0$, and
(\ref{U-on-z}) gives the linear action of $\U$ on $z_{j,P}$. 
To get the equality from (\ref{U-on-z}), one has to normalize 
only one component of $\Phi^*$.

Some elements of $\mbox{\sf det}({\bf L})$, corresponding 
to the border of the Newton polygon $\Pi$ of $\J(A,B)$, 
can be easily calculated.
Operator $d$ introduced is defined up to any integral of 
motion.
The simplest integrals are
\begin{equation}
\ds
j_1\;=\;\prod_P\;\u_{2,P}\;\u_{3,P}\;,\;\;\;\;
j_2\;=\;\prod_P\;\u_{1,P}\;\w_{3,P}^{-1}\;,\;\;\;
j_3\;=\;\prod_P\;\w_{1,P}\;\w_{2,P}\;,
\end{equation}
and the convenient choice of $d$ is
\begin{equation}\label{mu}
\ds d\;=\;
\prod_{P}\; \biggl( q^{1/2}\;
\u_{2,P}\*\u_{3,P}\*\w_{3,P}\biggr)^{-1}\;.
\end{equation}
$d$ can be absorbed into det,
\begin{equation}
\ds\J\;\;=\;\;\mbox{\sf det}\;({\bf L}^{(0)})\;,
\end{equation}
where
\begin{equation}\label{L0-matrix}
\ds
{\bf L}^{(0)}\;\;=\;\;
\left(\begin{array}{rcrcr}
\w_1^{} &,& q^{-1/2}\;\u_2^{-1} &,& q^{-1/2}\;\w_3^{-1}\\
&&&&\\
q^{1/2}\;\u_1^{} &,& T_a\;q^{1/2}\;\kappa_2\;\w_2^{} &,& 
T_b\;\u_3^{-1} \\
&&&&\\
T_a^{-1}\+ T_b^{-1}\;\kappa_1\;\u_1^{}\;\w_1^{} &,&
q^{1/2}\;\u_2^{-1}\;\w_2^{}\+T_b^{-1} &,&
T_a^{-1}\;\w_3^{-1}\;\u_3^{-1}\+\kappa_3
\end{array}\right)\;,
\end{equation}
Whole number of $\J_{\alpha,\beta}$ is
$3M^2+3M+1$, and there are $3M^2+1$
independent between them, and between these one can 
choose only $3M^2$
commutative, so $\J$ gives the complete set of integrals.
(As to whole number of summands in $\J$, 
e. g. for $M=2$
it is $1536\;=\;2^9\; 3$.)
The existence of $3\,M^2$ abelian integrals is the hypothesis
tested for small $M$-s.

All integrals corresponding to the boundary of domain $\Pi$, 
$|\alpha|=M$, $|\beta|=M$, $|\alpha+\beta|=M$,
are equivalent to the following $3M$ elements:
\begin{equation}\label{overline}
\ds\left\{\begin{array}{ccl}
\ds \overline{\u}_j & = & \ds
\prod_{\sigma}\;\;
\w^{-1}_{1,a^\sigma b^j P_0}\;
\w^{-1}_{2,a^\sigma b^j P_0}\;,\\
&&\\
\ds \overline{\v}_j & = & \ds
\prod_{\sigma}\;\;
\u^{}_{2,a^{j+\sigma} b^{-\sigma} P_0}\;
\u^{}_{3,a^{j+\sigma} b^{-\sigma} P_0}\;,\\
&&\\
\ds \overline{\w}_j & = & \ds
\prod_{\sigma}\;\;
\u^{}_{1,a^j b^{\sigma} P_0}\;
\w^{-1}_{3,a^j b^{\sigma} P_0}\;,
\end{array}\right.
\end{equation}
where $P_0$ is some frame of reference's point as previously.
Note,  $\overline{\v}_j$ are not $T_a,T_b$ -- invariant, 
but restoring
this invariance in any way, one obtains the invariants of $\U$.
Between $\overline{\w}_j,\overline{\u}_j,\overline{\v}_j$
one may choose $3M-1$ commutative elements.
Inner part of $\Pi$ gives $3M^2-3M+1$ highly complicated
independent integrals, which gives $g\;=\;3M^2-3M+1$ commutative
(up to (\ref{overline})) independent elements. Note, $g$ is the
formal genus of the curve $\J(A,B)=const$.

\subsection{Walks on the lattice and the integrals of motion}

Give now a geometrical interpretation of the integrals of motion.
This interpretation follows directly from the analysis of the determinant.
Every integral of motion is a sum of monomials associated with walks 
on the lattice such that all the walks have the same homotopy class
with respect to the torus on which the kagome lattice is defined.

It is useful to formulate the walks in terms of general
\underline{vertex} variables $\a$, $\b$, $\c$ and $\d$ as in Fig.
\ref{fig-abcd-currents}. Recall the shorter notation 
$W\;=\;\{\a,\b,\c,\d\}$ for the dynamical variables' set.
Consider the matrix ${\bf L}$ in this general case. Each row
in ${\bf L}$ corresponds to a vertex of the lattice,
and each column of ${\bf L}$ corresponds to a polygon 
(i.e. to a site) of the lattice. Thus $\mbox{\sf det} ({\bf L})$ 
consists on the monomials, each of them corresponds (up to a sign)
to a product of different $W_{j,P}$ such that:
\begin{itemize}
\item for any vertex $(j,P)$ \underline{only one} 
of $\a_{j,P}$, $\b_{j,P}$, $\c_{j,P}$, $\d_{j,P}$
is taken in this monomial, and
\item for any site \underline{only one} 
of surrounding $\a$, $...$, $\d$
is taken in this monomial.
\end{itemize}
Take the lattice and mark the places of the 
vertex variables $\a$, $...$, $\d$, corresponding to the monomial,
by the arrows, ingoing to the corresponding vertices. Thus
for any site and for any vertex we have painted only one arrow.

In order to get a purely invariant functional, we have to multiply
$\mbox{\sf det} ({\bf L})$ by an integrating monomial, in general case
this monomial is, for example, 
$\ds\prod_{P}\;\a_{1,P}^{-1}\;\d_{2,P}^{-1}\;\b_{3,P}^{-1}$.
This choice of the integrating multiplier corresponds to element
$d$ given by (\ref{mu}).
It is easy to see that this monomial has the same structure
as described above. But due to the power $-1$ we may interpret 
geometrically this monomial as the set of outgoing arrows.

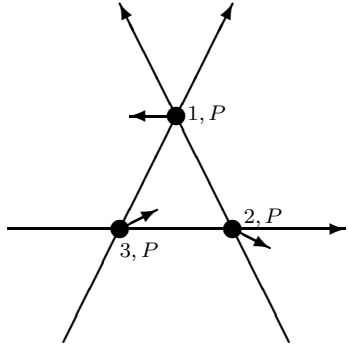
\begin{figure}
\begin{center}

\setlength{\unitlength}{0.25mm} 
\thicklines
\begin{picture}(450,200)
\put(125,0){
\begin{picture}(200,200)
\put(  10 ,  70 ){\vector(1,0){180}}
\put(  40 ,  10 ){\vector(1,2){90}}
\put( 160 ,  10 ){\vector(-1,2){90}}
\put(  70 ,  70 ){\circle*{10}}\put(70,55){\scriptsize $3,P$}
\put( 130 ,  70 ){\circle*{10}}\put(136,73){\scriptsize $2,P$}
\put( 100 , 130 ){\circle*{10}}\put(106,128){\scriptsize $1,P$}
\put(100,130){\vector(-1,0){25}}
\put(70,70){\vector(2,1){20}}
\put(130,70){\vector(2,-1){20}}
\end{picture}}
\end{picture}
\end{center}
\caption{Fixed outlets for the lattice walks}
\label{fig-outlet}
\end{figure}

The system of the outgoing arrows is thus fixed and shown in Fig.
\ref{fig-outlet} for each $\lgr$ -- type triangle of the lattice. 
For the system of the outgoing arrows and 
any system of ingoing arrows the following is valid:
\begin{itemize}
\item for any site there exist exactly one outgoing arrow and
exactly one ingoing arrow, and they may touch the same vertex, and
\item for any vertex there exist exactly one outgoing arrow
and exactly one ingoing arrow, and they may belong to the same
site.
\end{itemize}
Hence there is the unique way to connect all the arrows inside each
site so that a walk appears.

So, the walks we consider, obey the following demands:
\begin{itemize}
\item the system of outlets of the walk is fixed and given by Fig.
\ref{fig-outlet},
\item the walk visits any site only once,
\item the walk must visit all the sites and
\item the walk must visit all the vertices.
\end{itemize}
For any walk ${\cal W}$ let $\sigma({\cal W})$ be the 
number of the components of the connectedness
(i.e. the number simply connected subwalks).

Let now walk ${\cal W}$ belongs to a given homotopy class 
$\alpha\;{\cal A}\;+\;\beta\;{\cal B}$
of the torus, where cycle ${\cal A}$ corresponds to $T_a^M$ and 
cycle ${\cal B}$ corresponds to $T_b^M$, and denote such walk as 
${\cal W}_{\alpha,\beta}$.

To a walk given assign a monomial according to the following rules:
let the walk pass through vertex $(j,P)$ so that the walk ingoes the vertex
form the side $\x$ $\in$ $W_{j,P}$,
and outgoes the vertex from the side 
$\y$ $\in$ $W_{j,P}$. 
Then the multiplier
corresponding to $(j,P)$ is $\ds\x\*\y^{-1}$. The monomial $\J_{\cal W}$
is the product of such multipliers corresponding to all
the vertices. Thus the reader may see that each monomial we construct
gains the structure of an element of ${\cal B}^\prime_P$, described
in section ``Auxiliary linear problem'', subsection ``General approach'':
monomial $\J_{\cal W}$,
\begin{equation}
\ds
\J_{\cal W}\;=\;
\cdots\;\;\x\*\y^{-1}\*\x^\prime\*\y^{\prime-1}\;\;\cdots\;,
\end{equation}
$\x$ and $\y$ are assigned to a same vertex, so $\x\*\y^{-1}$
does not contain the vertex projective ambiguity, and $\y$ and $\x^\prime$
belong to a same site, so $\y^{-1}\*\x^\prime$ does not contain the
site ambiguity. Finally we have to provide the projective invariance 
of $\J_{\cal W}$ with respect to the start and end points
of each simply connected subwalk. In our case of the local Weyl algebrae
this invariance is obvious, because of elements $\x\*\y{-1}$ for different
vertices commute.

With the structure of the walks introduced, the simple analysis of the
determinant gives immediately
\begin{equation}
\ds
\J_{\alpha,\beta}\;\;=\;\;
\sum_{\ds\mbox{all}\;\;{\cal W}_{\alpha,\beta}}
\;\;\; (-)^{\ds\sigma({\cal W}_{\alpha,\beta})}
\;\*\; \J_{\ds{\cal W}_{\alpha,\beta}}\;,
\end{equation}
where the sum is taken over all the walks of the homotopy
class $\alpha\;{\cal A}\;+\;\beta\;{\cal B}$ given
and the system of the outlets of the walks fixed.

\subsection{Monodromy operator}

Consider now another interpretation of the two 
dimensional kagome lattice.

Let now to each vertex of the lattice the local 
$\L$-operator is assigned.
Instead of $\omega$ (\ref{omega}) 
in the definition of $\L$, use
\begin{equation}
\ds \x'\;=\;\C^{-1}\*\x\;,\;\;\;\;\;
\y'\;=\;\C^{}\*\y\;,
\end{equation}
where
\begin{equation}
\ds\C\;=\;\C(\x^{-1}\u , \y^{-1}\w)\;
\end{equation}
and
\begin{equation}
\ds
\C(\u,\w)\;=\;\u^{-1}\;-\;q^{1/2}\;\u^{-1}\w\;+\;
\kappa\;\w\;,
\end{equation}

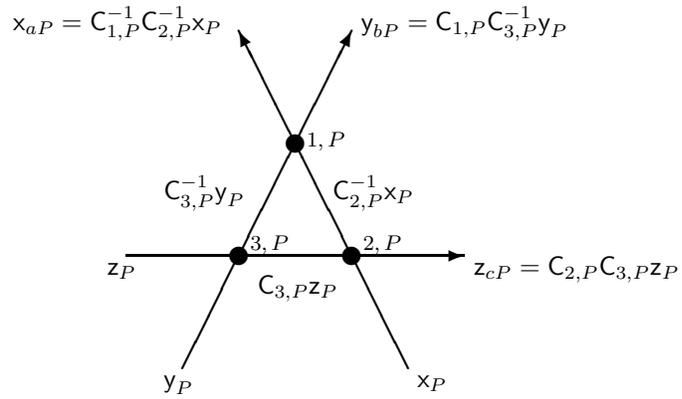
\begin{figure}
\begin{center}

\setlength{\unitlength}{0.25mm} 
\thicklines
\begin{picture}(450,200)
\put(125,0){
\begin{picture}(200,200)
\put(  10 ,  70 ){\vector(1,0){180}}
\put(  40 ,  10 ){\vector(1,2){90}}
\put( 160 ,  10 ){\vector(-1,2){90}}
\put(  70 ,  70 ){\circle*{10}}\put(76,73){\scriptsize $3,P$}
\put( 130 ,  70 ){\circle*{10}}\put(136,73){\scriptsize $2,P$}
\put( 100 , 130 ){\circle*{10}}\put(106,128){\scriptsize $1,P$}
\put(0,60){$\z_P$}
\put(80,50){$\C_{3,P}\z_P$}
\put(195,60){$\z_{cP}=\C_{2,P}\C_{3,P}\z_P$}
\put(30,0){$\y_P$}
\put(30,100){$\C_{3,P}^{-1}\y_P$}
\put(135,190){$\y_{bP}=\C_{1,P}\C_{3,P}^{-1}\y_P$}
\put(165,0){$\x_P$}
\put(120,100){$\C_{2,P}^{-1}\x_P$}
\put(-50,190){$\x_{aP}=\C_{1,P}^{-1}\C_{2,P}^{-1}\x_P$}
\end{picture}}
\end{picture}
\end{center}
\caption{Edge variables of the triangle}
\label{fig-edge-triangle}
\end{figure}

For the $\lgr$-type triangle $P$ 
let the in - edge variables be $\x_P$, $\y_P$,
$\z_P$ so that out edge variables are  $\x_{aP}$, $\y_{bP}$
and $\z_{cP}$, $c\;=\;a^{-1}\, b$. 
These notations are shown in Fig.
\ref{fig-edge-triangle}. Surely LYBE for $\L$-operators means that
for $\x_P$, $\y_P$, $\z_P$ given, $\x_{aP}$, $\y_{bP}$ and
$\z_{cP}$ are the same for the right hand side YBE graph $\rgr$.

Consider now the whole toroidal kagome lattice. 
We are going to assign
$\x$, $\y$ and $\z$ to some minimal set of the 
edges so that the variables
of all other edges can be restored via functions $\C_{j,P}$.

\begin{figure}
\begin{center}

\setlength{\unitlength}{0.25mm} 
\thicklines
\begin{picture}(400,400)
\put(0,0){
\begin{picture}(400,400)
\multiput(100,400)(0,-20){10}{\line(0,-1){15}}
\multiput(100,200)(20,-20){10}{\line(1,-1){15}}
\multiput(300,400)(0,-20){10}{\line(0,-1){15}}
\multiput(300,200)(20,-20){5}{\line(1,-1){15}}
\multiput(0,100)(20,-20){5}{\line(1,-1){15}}
\put(100,220){\vector(1,0){200}}
\put(100,200){\vector(1,0){200}}
\put(120,180){\vector(1,0){200}}
\put(85,215){$\y_2$}\put(85,195){$\y_1$}\put(105,175){$\y_0$}
\put(180,120){\vector(0,1){280}}\put(180,0){\vector(0,1){100}}
\put(200,100){\vector(0,1){300}}\put(200,0){\vector(0,1){80}}
\put(220,80){\vector(0,1){320}}\put(220,0){\vector(0,1){60}}
\put(175,105){$\x_0$}\put(195,85){$\x_1$}\put(215,65){$\x_2$}
\put(100,320){\vector(1,-1){300}}\put(20,400){\vector(1,-1){60}}
\put(100,300){\vector(1,-1){300}}\put(0,400){\vector(1,-1){80}}
\put(100,280){\vector(1,-1){280}}\put(0,380){\vector(1,-1){80}}
\put(0,20){\vector(1,-1){20}}\put(380,400){\vector(1,-1){20}}
\put(85,329){$\z_2$}\put(85,309){$\z_1$}\put(85,289){$\z_0$}
\put(0,0){\line(1,0){400}}\put(0,0){\line(0,1){400}}
\put(0,400){\line(1,0){400}}\put(400,0){\line(0,1){400}}
\end{picture}}
\end{picture}
\end{center}
\caption{The string (dashed) on the toroidal kagome lattice.}
\label{fig-monodromy}
\end{figure}
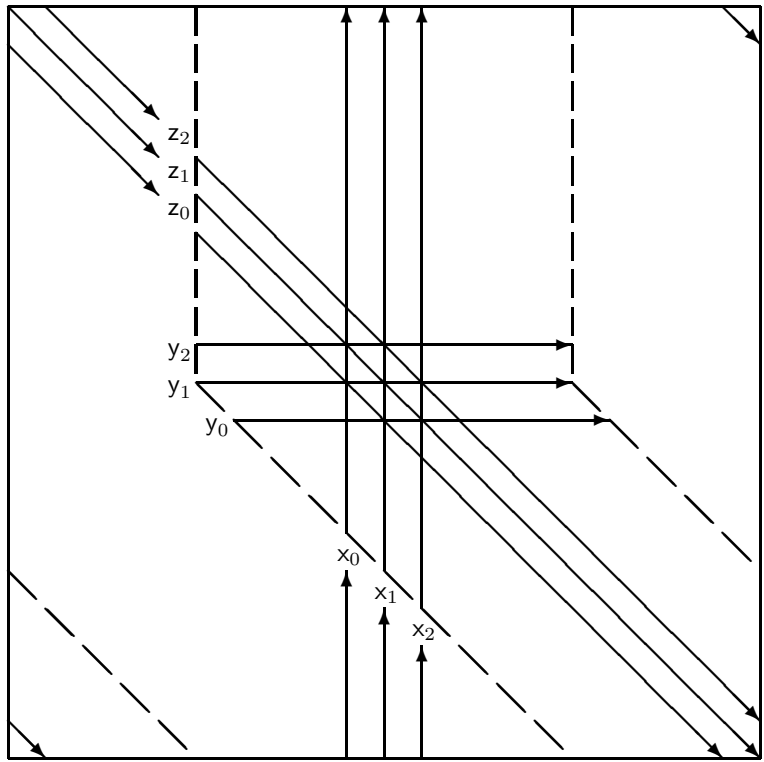

To do this, cut the torus along some line, 
shown as the dashed line
in Fig. \ref{fig-monodromy}.
Call this line `the string'.
The edge variables along the string we'll denote as
$\x_j$, $\y_j$ and $\z_j$.
It is useful to draw
the string so that it intersects all $\x$ and $\z$ lines once,
and $\y$-lines twice (i.e. the homotopy class of the string
is $\pm(2\;{\cal A}-{\cal B})$, an orientation of the string 
and so a sign are unessential)
Note, $\x_j$, $\y_j$ and $\z_j$ introduced we assign to
the edges which are right-touched to the string.

Now switch on the $\L$-operator game with the edge variables.
We interpret it as the shift of the string.
We enumerate the lines so that the triangle 
$P\;=\;a^{j}\;b^{i}\;P_0$
is surrounded by the lines $\x_{i}$, $\y_{j}$ 
and $\z_{i+j}$, so as
the $\L$ -- operators are
\begin{equation}\label{L-on-lattice}
\ds
\L_{\x_i,\y_j}(\{\u,\w\}_{1,a^{j}\,b^{i}\,P_0})\;,\;\;\;\;
\L_{\x_i,\z_k}(\{\u,\w\}_{2,a^{k-i}\,b^{i}\,P_0})\;,\;\;\;\;
\L_{\y_j,\z_k}(\{\u,\w\}_{3,a^{j}\,b^{k-i}\,P_0})\;.
\end{equation}
The $\L$-operator game allows us to restore all the edge variables
for the lattice, including the left-touched to the original string
variables $\x_j'$, $\y_j'$, $\z_j'$.

Thus for given variables from the right side of the string
we obtain the analogous values from the left side of the string
as functionals of the given variables. Thus the map corresponding
to the kagome lattice and the string chosen appears:
\begin{equation}
\ds T\;(\L)\;:\;\{\x_i,\y_j,\z_k\}\;\;\mapsto\;\;
\{\x'_i,\y'_j,\z'_k\}\;.
\end{equation}

As the operator, $T(\L)$ is ordered product of all $\L$
(\ref{L-on-lattice}). Define $A<B$ if the ordered product 
of $A$ and $B$ is $A\* B$. Then in $T(\L)$
\begin{eqnarray}
&\ds\L_{\y_j,\z_{i+j}}\;<
\;\L_{\x_i,\z_{i+j}}\;<
\;\L_{\x_i,\y_j}\;,&\\
&\ds\L_{\x_i,\y_j}\;<
\;\L_{\x_i,\z_{i+j+1}}\;<
\;\L_{\y_j,\z_{i+j+1}}\;.&
\end{eqnarray}
These relations are enough to restore $T(\L)$.
Operator $T(\L)$ resembles the monodromy matrix in $2D$. 
The difference is that instead of the distinguished point 
in $2D$ monodromy matrix
(i.e. the point where the transfer matrix is torn),
in $3D$ we have the distinguished string.

Now, what should stand for a ``trace'' of $T(\L)$. 
Consider the system
\begin{equation}\label{trace}
\x_j'\;=\;\x_j\;,\;\;\;\;\y_j'\;=\;\y_j\;,\;\;\;\;\z_j'\;=\;\z_j
\end{equation}
on some left module element $\phi^*_0$. Here are $3\,M$ equations,
$3\,M-1$ from them are independent due to
\begin{equation}
\ds\prod_j\;\;\x_j\;\y_j\;\z_j\;=
\;\prod_j\;\;\x_j'\;\y_j'\;\z_j'\;,
\end{equation}
where it is implied that all $\x_{P},\y_{P},\z_{P}$ $\forall P$ are  
commutative, this is the consequence of the commutability of primary
$\x_j,\y_j,\z_j$.
Then solve $3\,M-2$ equations of (\ref{trace}) 
leaving two variables,
up to unessential signs and powers of $q$ :
\begin{equation}
\ds
A\;=\;\prod_{j}\;\;q\;\y_j^{}\;\z_j^{}\;,\;\;\;\;
B\;=\;\prod_{j}\;\;\x_j^{-1}\;\z_j^{}\;.
\end{equation}
A single equation rests for $A$ and $B$, and amusingly 
this equation coincides with
the quantum determinant relation $\phi_0^*\*\J(A,B)=0$.
So in this sense $\J(A,B)\;=\;0$ is the trace of the 
monodromy operator.

Note however, $\J(A,B)$ is {\bf the invariant curve},
this was established in the previous section,
so it is not necessary to consider $\phi^*_0\*\J(A,B)\;=\;0$.
The actual problem for the further investigations is to 
diagonalize $\J(A,B)$ for $A$ and $B$ arbitrary.


\section{Discussion}

Conclude this paper by an overview of the problems to be solved
and the aims to be reached.
The approach proposed gives a way for their solution.

First, mention the problems of the classification of the map
\begin{equation}
\ds
\R\;\;\;:\;\;\;\{\a_j,\b_j,\c_j,\d_j\}\;\;\mapsto\;\;
\{\a_j',\b_j',\c_j',\d_j'\}\;,
\;\;\;\;j=1,2,3, 
\end{equation}
in general.
The aim is to classify all conserving
symplectic structures of the body ${\cal B}$.
We have discussed only the local case, when the
variables, assigned to different vertices commute, and the scalars
(spectral parameters) are conserved. We suppose, such
case is not unique, and there are another ways to remove the projective
ambiguity. The simplest case to be investigated is to consider all
the variables $\a,\b,\c,\d$ for each vertex as matrices with,
for example, non-commutative entries, but with this entries commutative
for any two vertices. The matrix structure may be common for all
vertices, and thus we would have no commutation between different
vertices in general. Another simple possibility is another
kind of locality, the case when the dynamical variables
commute while do not belong to a same site.

Note, once our locality is imposed, the Weyl structure appears 
immediately. Thus the Weyl algebra is the consequence of the locality
technically, but a principal origin of the Weyl algebra is 
mysterious. 

Next fundamental problem is the quantization of 
Korepanov's matrix model mentioned above. The conservation of
complete algebra ${\cal X}$, (\ref{algebra},\ref{z-eq}), 
means that we can not use
(\ref{ad-bc}) to fix all the ambiguity of Korepanov's equation.
Analysis of (\ref{KE}) {\em plus} some extra (but natural) symmetry
conditions allows to fix the functional map $r$ : $X_j^{}\mapsto X_j'$ up
to one unknown function of three variables. 
The problem of the Tetrahedron equation for these $r$ is open.
All these are a subject of a separate paper.

Pure technical problem to be mentioned is the investigation 
of $q$ - hypergeometrical function $\sigma$, 
eqs. (\ref{sigma-psi},\ref{sigma-chi}).

Another interesting thing is functional $\L$ -- operators and
LYBE related to them. The map given by $\L$,
eqs. (\ref{omega},\ref{L-op}), is a bi-rational one.
Note, the case of linear $\L$ coincides with Korepanov's $X$.
Thus the rational
case of it as well as the general case of the bi-rational transformation
have good perspectives for the investigation.

The main set of problems for further investigations is connected
with the integrals of $\U$. $\J(A,B)$ seems to be not
constructive. The aim is to calculate the spectrum of it,
and to calculate $\U$ as a function of its integrals.
Possible approach is functional equations for the integrals of motion,
that should follow from the determinant or topological 
representation of $\J$.

Another possibility is a way resembling the Bethe ansatz
in 2D should exist in 3D, i.e. a way of a triangulation
of $\U$ with a help of some artificial operators.
If such way exists, it must based
on the linear problem derived.

\noindent
{\bf Acknowledgements} I would like to thank sincerely
Rinat Kashaev, Igor Korepanov
and Alexey Isaev for their interest to this work and
many fruitful discussions. Many thanks also for
Yu. Stroganov, G. Pronko, V. Mangazeev and H. Boos.

The work was partially almost supported by the RFBR grant 
No. 98-01-00070.


\end{document}